\documentclass[submission,copyright,creativecommons]{eptcs}

\usepackage{holtexbasic,proof}
\usepackage{stmaryrd}

\usepackage[utf8]{inputenc}
\usepackage[T1]{fontenc}
\usepackage[english]{babel}
\providecommand{\event}{LFMTP 2020} 
\usepackage{breakurl}             
\usepackage{underscore}           
\usepackage{amsmath,amssymb,stmaryrd,amsthm}
\usepackage{graphicx}
\usepackage{mathpartir}
\usepackage{todonotes}

\usepackage{microtype}
\usepackage{hyperref}
\usepackage[capitalise]{cleveref}
\usepackage{xspace}

\title{Mechanisation of Model-theoretic Conservative Extension\\ for HOL with Ad-hoc Overloading}
\author{
Arve Gengelbach
\institute{Uppsala University, Uppsala, Sweden}
\email{arve.gengelbach@it.uu.se}
\and
Johannes {\AA}man Pohjola
\institute{CSIRO's Data61, Sydney, Australia}
\institute{University of New South Wales, Sydney, Australia}
\email{johannes.amanpohjola@data61.csiro.au}
\and
Tjark Weber
\institute{Uppsala University, Uppsala, Sweden}
\email{tjark.weber@it.uu.se}
}
\def\titlerunning{Mechanisation of Model-theoretic Conservative Extension for HOL}
\def\authorrunning{A. Gengelbach, J. {\AA}man Pohjola \& T. Weber}

\newcommand{\bool}{\ensuremath\mathsf{Bool}}

\newcommand{\GType}{\ensuremath\mathsf{GType}}

\newcommand{\GCInst}{\ensuremath{\mathsf{GCInst}}}
\newcommand{\rep}{\ensuremath{\mathsf{rep}}}
\newcommand{\abs}{\ensuremath{\mathsf{abs}}}
\newcommand{\dep}{\ensuremath{\rightsquigarrow}} 

\newcommand{\types}{{}^\bullet{}}
\newcommand{\consts}{{}^\circ{}}

\newcommand{\etal}{et al.\xspace}
\newcommand{\wrt}{w.\,r.\,t.\@\xspace}
\newcommand{\ie}{i.\,e.\@\xspace}

\newcommand{\eg}{e.\,g.\@\xspace}

\renewcommand{\phi}{\ensuremath\varphi}
\newcommand{\N}{\ensuremath\mathbb{N}}

\newcommand\restr[2]{{
  \left.\kern-\nulldelimiterspace 
  #1 
  \vphantom{\big|} 
  \right|_{#2} 
  }}

\makeatletter
\def\moverlay{\mathpalette\mov@rlay}
\def\mov@rlay#1#2{\leavevmode\vtop{%
   \baselineskip\z@skip \lineskiplimit-\maxdimen
   \ialign{\hfil$\m@th#1##$\hfil\cr#2\crcr}}}
\newcommand{\charfusion}[3][\mathord]{
    #1{\ifx#1\mathop\vphantom{#2}\fi
        \mathpalette\mov@rlay{#2\cr#3}
      }
    \ifx#1\mathop\expandafter\displaylimits\fi}
\makeatother

\newcommand{\dependencypair}[3]{\ensuremath{#2\;\mathrel{\rightsquigarrow_{#1}}\;#3}}

\newcommand{\dependencyrtcpair}[3]{\ensuremath{#2\;{({\rightsquigarrow_{#1}}^{\downarrow})}^*\;#3}}
\newcommand{\HOLTokenSupCirc}{\ensuremath{{}^\circ}}
\newcommand{\HOLTokenSupBullet}{\ensuremath{{}^\bullet}}

\renewcommand{\HOLTokenTurnstile}{\ensuremath{\vdash\!\!}}
\renewcommand{\HOLConst}[1]{\textsf{\small #1}}

\renewcommand{\HOLSymConst}[1]{\HOLConst{#1}}
\renewcommand{\HOLTyOp}[1]{\HOLConst{#1}}
\renewcommand{\HOLinline}[1]{\textsf{\ensuremath{#1}}}
\renewcommand{\HOLKeyword}[1]{\mathsf{#1}}
\renewcommand{\HOLTokenBar}{\ensuremath{\mathtt{|}}}

\usepackage{environ}
\NewEnviron{holthmenv}{%
  \[
  \scalebox{1.0}{\begin{array}[t]{l}
  \BODY
  \end{array}}
  \]}

\NewEnviron{holthmenvl}{%
\begin{equation}\begin{aligned}
  \scalebox{0.9}{\begin{array}[t]{l}\BODY\end{array}}
\end{aligned}\end{equation}}

\begin{document}
\maketitle

\begin{abstract}
Definitions of new symbols merely abbreviate expressions in logical frameworks, and
no new facts (regarding previously defined symbols) should hold because of a new definition.
In Isabelle/HOL, definable symbols are types and constants.  The latter may be
ad-hoc overloaded, \ie have different definitions for non-overlapping
types.
We prove that symbols that are independent of a new definition may keep their interpretation in a model extension.
This work revises our earlier notion of model-theoretic conservative extension and generalises an earlier model construction.
We obtain consistency of theories of definitions in higher-order logic (HOL) with ad-hoc overloading as a corollary.
Our results are mechanised in the HOL4 theorem prover.
\end{abstract}

\section{Introduction}\label{sec:introduction}
Isabelle/HOL enriches higher-order logic with ad-hoc overloading.
While other theorem provers of the HOL family support overloaded syntax through enhancements of parsing and pretty printing, in Isabelle/HOL overloading is a feature of the logic.
The user-defined symbols are types and constants, and in Isabelle/HOL the latter may have multiple definitions for non-overlapping types.
For instance,~$+_{\alpha\to\alpha\to\alpha}$ is an overloaded constant with different definitions for different type instances of commutative monoids such as the natural numbers or the integers.

Overloaded definitions need further care as the defined symbols may be used prior to their definition, which if treated improperly may lead to cyclic definitions, \ie unfolding of definitions might not terminate.

For a logic to be useful it should have unprovable statements.
A logic is \emph{consistent} if a contradiction cannot be deduced from any of its theories.
HOL with user-defined types and constants without overloading is consistent.  This can be proved by an argument based on \emph{standard semantics}~\cite{pitts1993}, where Booleans, function types and the equality constant are interpreted as expected, and type variables are interpreted as elements of a fixed universe of sets.
The consistency story for HOL with overloading is a long one~\cite{DBLP:conf/tphol/Wenzel97,DBLP:conf/rta/Obua06,DBLP:journals/jar/KuncarP19a,LPAR23:AmPoGen20}.
Åman Pohjola and Gengelbach~\cite{LPAR23:AmPoGen20} prove HOL with overloading consistent in a machine-checked proof, by constructing models of theories of definitions through a construction that originates from Kunčar and Popescu~\cite{DBLP:journals/jar/KuncarP19a}.
Kunčar and Popescu introduce a \emph{dependency relation} between the symbols of a theory to track dependencies of defined symbols on their definiens.  Under an additional syntactic restriction on overloading~\cite{DBLP:conf/cpp/Kuncar15}, they construct a model for any (finite) theory of definitions for which the dependency relation is terminating.

Apart from consistency, a definitional mechanism should be \emph{model-theoretically conservative}:
any model of a theory can be extended to a model of the extended theory with new definitions, keeping interpretations of formulae that are independent of the new symbols intact.
Informally, at least symbols that are independent of a theory extension may keep their interpretation in a model extension.

For HOL without overloading, model-theoretic conservativity holds unconditionally~\cite{KuArMyOw14}. With overloading,
model-theoretic conservativity holds for symbols that are independent of new definitions, as Gengelbach and Weber prove~\cite{DBLP:journals/entcs/GengelbachW18}.
However, their proof was based on inherited wrong assumptions from Kunčar and Popescu~\cite{DBLP:journals/jar/KuncarP19a}, that Åman Pohjola and Gengelbach in their mechanised model construction~\cite{LPAR23:AmPoGen20} uncover and correct.
Additionally, the mechanisation supports theory extension by the more expressive constant specification~\cite{Art14}, which is a definitional mechanism also used in the theorem provers ProofPower and HOL4 to simultaneously introduce several new constants that satisfy some property.

This paper joins these two lines of work in mechanising that the definitional mechanisms of types and overloaded constants are model-theoretically conservative.
The result holds for models that interpret constants introduced by constant specification equal to their witnesses and replaces the earlier monolithic model construction with an iterative one.
An interpretation of this result is that the definitional mechanisms of Isabelle/HOL are semantically speaking robustly designed: at least symbols that are independent of an update may keep their interpretation in a model extension.

Even more generally, the syntactic counterpart of model-theoretic conservativity, \emph{proof-theoretic (syntactic) conservativity} shall hold~\cite{DBLP:conf/tphol/Wenzel97}.
Informally, a definitional mechanism is \emph{proof-theoretically conservative} if the definitional extension entails no new properties, except those that depend on the symbols which the extension introduces.

For Isabelle/HOL conservativity has been studied in an absolute manner, \ie any definitional theory is a conservative extension of \emph{initial HOL}~\cite{DBLP:journals/pacmpl/Kuncar018}, the theory of Booleans with Hilbert-choice and infinity~axiom.
Gengelbach and Weber~\cite{GengelbachW20} prove conservativity of any definitional extension above initial HOL, by translating a model-theoretic conservativity for a generalised semantics into its syntactic counterpart.
As their semantics are similar to ours, this paper adds to the reliability of their result.

We describe the syntax and (lazy ground) semantics of HOL with ad-hoc overloading in \cref{sec:background}.
Subsequently, in \cref{sec:independentfragment} we recapitulate the \emph{independent fragment} as the part of a theory that is independent of an extension by a new definition.
This fragment is crucial in the iterative model construction, \ie model-theoretic conservativity in \cref{sec:conservativity}.
We discuss related work in \cref{sec:relatedwork}.
The definitions and theorems in this paper are formalised in the HOL4 theorem prover as part of the CakeML project.\footnote{\url{https://code.cakeml.org/tree/master/candle/overloading/semantics/}}

\paragraph*{Contributions}
We make the following contributions
\begin{itemize}
  \item
    We adapt and formalise the previously introduced \emph{independent fragment}~\cite{DBLP:journals/entcs/GengelbachW18} (\ie a theory's syntax fragment that is independent of a theory update) to support a more general definitional mechanism for constants:
    \emph{constant specification}~\cite{Art14}.
  \item
    We use the independent fragment to prove a notion of model-theoretic conservativity~\cite{DBLP:journals/entcs/GengelbachW18} in a new setting for the \emph{lazy ground semantics}~\cite{LPAR23:AmPoGen20}, which delays type variable instantiation and does not instantiate the type of term variables.
    To the best of our knowledge, this is the first mechanised
    conservativity result for a logic with overloaded definitions.
  \item
    Our work generalises and replaces the earlier monolithic model construction of Åman Pohjola and Gengelbach~\cite{LPAR23:AmPoGen20}, and obtains consistency of HOL with ad-hoc overloading as a corollary.
\end{itemize}

\section{Background}\label{sec:background}
In this section we introduce the syntax and semantics of HOL with ad-hoc overloading, which we inherit from the earlier work of Åman Pohjola and Gengelbach~\cite{LPAR23:AmPoGen20}.
Their formalisation makes use of infrastructure from the formalisation of HOL Light (without overloading) by Kumar \etal~\cite{DBLP:journals/jar/KumarAMO16}, and the theoretical work on the consistency of HOL with ad-hoc overloading by Kunčar and Popescu~\cite{DBLP:journals/jar/KuncarP19a}.

\subsection{Types and terms}\label{sec:syntax}

\paragraph*{Types}
Types, described by the grammar \HOLTyOp{type}\;=\;\HOLConst{Tyvar}\;\HOLTyOp{string}\;\HOLTokenBar{}\;\HOLConst{Tyapp}\;\HOLTyOp{string}\;(\HOLTyOp{type}\;\HOLTyOp{list}), are rank-1 polymorphic.
Type variables \HOLinline{\HOLConst{Tyvar}} can be instantiated by a type substitution (ranged
over by \HOLinline{\HOLFreeVar{\varTheta{}}}), which extends homomorphically to type constructors
\HOLinline{\HOLConst{Tyapp}}.
For a type \HOLinline{\HOLFreeVar{ty}} and a type substitution \HOLinline{\HOLFreeVar{\varTheta{}}}, we call \HOLinline{\HOLFreeVar{\varTheta{}}\;\HOLFreeVar{ty}}
a \emph{(type) instance} of \HOLinline{\HOLFreeVar{ty}}, denoted by
\HOLinline{\HOLFreeVar{ty}\;\HOLSymConst{\ensuremath{\geq}}\;\HOLFreeVar{\varTheta{}}\;\HOLFreeVar{ty}}.
Two types \HOLinline{\ensuremath{\HOLFreeVar{ty}\sb{\mathrm{1}}}} and \HOLinline{\ensuremath{\HOLFreeVar{ty}\sb{\mathrm{2}}}} are \emph{orthogonal}, denoted by
\HOLinline{\ensuremath{\HOLFreeVar{ty}\sb{\mathrm{1}}}\;\HOLSymConst{\ensuremath{\#}}\;\ensuremath{\HOLFreeVar{ty}\sb{\mathrm{2}}}}, if they have no common instance. \emph{Ground} types are
those that contain no type variables, hence remain unchanged under any type
substitution. A type substitution is \emph{ground} if it maps every type to
a ground type. As ground types only have trivial instances, any two ground
types are either equal or~orthogonal.

\paragraph*{Terms}
Terms are simply typed $\lambda$-expressions, described by the grammar
\begin{holthmenv}
\HOLTyOp{term}\;=\;\HOLConst{Var}\;\HOLTyOp{string}\;\HOLTyOp{type}\;\HOLTokenBar{}\;\HOLConst{Const}\;\HOLTyOp{string}\;\HOLTyOp{type}\;\HOLTokenBar{}\;\HOLConst{Comb}\;\HOLTyOp{term}\;\HOLTyOp{term}\;\HOLTokenBar{}\;\HOLConst{Abs}\;\HOLTyOp{term}\;\HOLTyOp{term}
\end{holthmenv}
We only consider well-formed terms, that is,
$\lambda$-abstractions must be of the form \HOLinline{\HOLConst{Abs}\;(\HOLConst{Var}\;\HOLFreeVar{x}\;\HOLFreeVar{ty})\;\HOLFreeVar{t}},
\ie have a term variable as first argument representing the binder.
A closed term \HOLinline{\HOLFreeVar{t}}, denoted \HOLinline{\HOLConst{closed}\;\HOLFreeVar{t}}, contains only bound term variables.
A term is \HOLinline{\HOLConst{welltyped}} if it has a type by the following rules (wherein $\to$ abbreviates the later introduced function type):
\begin{mathpar}
\infer{\HOLinline{\HOLConst{Var}\;\HOLFreeVar{n}\;\HOLFreeVar{ty}\;\HOLConst{has_type}\;\HOLFreeVar{ty}}}{}
\and
\infer{\HOLinline{\HOLConst{Const}\;\HOLFreeVar{n}\;\HOLFreeVar{ty}\;\HOLConst{has_type}\;\HOLFreeVar{ty}}}{}\\
\infer{\HOLinline{\HOLConst{Comb}\;\HOLFreeVar{s}\;\HOLFreeVar{t}\;\HOLConst{has_type}\;\HOLFreeVar{rty}}}{\HOLinline{\HOLFreeVar{s}\;\HOLConst{has_type}\;(\HOLFreeVar{dty}\HOLSymConst{\ensuremath{\;\to{}}}\HOLFreeVar{rty})}&\HOLinline{\HOLFreeVar{t}\;\HOLConst{has_type}\;\HOLFreeVar{dty}}}
\and
\infer{\HOLinline{\HOLConst{Abs}\;(\HOLConst{Var}\;\HOLFreeVar{n}\;\HOLFreeVar{dty})\;\HOLFreeVar{t}\;\HOLConst{has_type}\;(\HOLFreeVar{dty}\HOLSymConst{\ensuremath{\;\to{}}}\HOLFreeVar{rty})}}{\HOLinline{\HOLFreeVar{t}\;\HOLConst{has_type}\;\HOLFreeVar{rty}}}
\end{mathpar}
A well-typed term \HOLinline{\HOLFreeVar{tm}} has a unique type which we denote \HOLinline{\HOLConst{typeof}\;\HOLFreeVar{tm}}.
Applying a type substitution \HOLinline{\HOLFreeVar{\varTheta{}}} to a term means to apply \HOLinline{\HOLFreeVar{\varTheta{}}} to the
types within, \eg \HOLinline{\HOLConst{\!\!}\;\HOLFreeVar{\varTheta{}}\;(\HOLConst{Const}\;\HOLFreeVar{c}\;\HOLFreeVar{ty})\;\HOLSymConst{=}\;\HOLConst{Const}\;\HOLFreeVar{c}\;(\HOLConst{\!\!}\;\HOLFreeVar{\varTheta{}}\;\HOLFreeVar{ty})}
(which we call \emph{constant instance}) for a constant \HOLinline{\HOLConst{Const}\;\HOLFreeVar{c}\;\HOLFreeVar{ty}}.
Orthogonality extends from types to constant instances:
\begin{holthmenv}
\HOLConst{Const}\;\HOLFreeVar{c}\;\ensuremath{\HOLFreeVar{ty}\sb{\mathrm{1}}}\;\HOLSymConst{\ensuremath{\#}}\;\HOLConst{Const}\;\HOLFreeVar{d}\;\ensuremath{\HOLFreeVar{ty}\sb{\mathrm{2}}}\;\HOLTokenDefEquality{}\;\HOLFreeVar{c}\;\HOLSymConst{\HOLTokenNotEqual{}}\;\HOLFreeVar{d}\;\HOLSymConst{\HOLTokenDisj{}}\;\ensuremath{\HOLFreeVar{ty}\sb{\mathrm{1}}}\;\HOLSymConst{\ensuremath{\#}}\;\ensuremath{\HOLFreeVar{ty}\sb{\mathrm{2}}}.
\end{holthmenv}

A user may introduce (non-built-in) types and constants by theory extension, as described in \cref{sec:theoryextensions}.
For types and constants we generally say \emph{symbols}.

\paragraph*{Built-ins} We abbreviate
\[\small
\begin{array}{l@{\quad}r@{\quad}l}
  \HOLinline{\HOLConst{Bool}}&  \mbox{for} & \HOLinline{\HOLConst{Tyapp}\;\HOLStringLitDG{bool}\;\HOLConst{[\,]}}\\
  \HOLinline{\HOLFreeVar{x}\HOLSymConst{\ensuremath{\;\to{}}}\HOLFreeVar{y}} & \mbox{for} & \HOLinline{\HOLConst{Tyapp}\;\HOLStringLitDG{fun}\;[\HOLFreeVar{x};\;\HOLFreeVar{y}]}\\
  \HOLinline{\HOLConst{Equal}\;\HOLFreeVar{ty}} & \mbox{for} & \HOLinline{\HOLConst{Const}\;\HOLStringLitDG{=}\;(\HOLFreeVar{ty}\HOLSymConst{\ensuremath{\;\to{}}}\HOLFreeVar{ty}\HOLSymConst{\ensuremath{\;\to{}}}\HOLConst{Bool})}\\
  \HOLinline{\HOLFreeVar{s}\;\HOLSymConst{===}\;\HOLFreeVar{t}} & \mbox{for} & \HOLinline{\HOLConst{Comb}\;(\HOLConst{Comb}\;(\HOLConst{Equal}\;(\HOLConst{typeof}\;\HOLFreeVar{s}))\;\HOLFreeVar{s})\;\HOLFreeVar{t}}
\end{array}
\]
Any type with any of these type constructors at the top level is \emph{built-in}, as is the constant $\mathsf{Equal}$.
These are the only symbols which are not user-defined.
A \emph{formula} is a term of type \HOLinline{\HOLConst{Bool}}.

For a set of types \HOLinline{\HOLFreeVar{tys}} we consider its \emph{built-in closure}, written \HOLinline{\HOLConst{builtin_closure}\;\HOLFreeVar{tys}}:
\begin{mathpar}
\infer{\HOLinline{\HOLConst{Bool}\;\HOLSymConst{\HOLTokenIn{}}\;\HOLConst{builtin_closure}\;\HOLFreeVar{tys}}}{} \and
\infer{\HOLinline{\HOLFreeVar{ty}\;\HOLSymConst{\HOLTokenIn{}}\;\HOLConst{builtin_closure}\;\HOLFreeVar{tys}}}{\HOLinline{\HOLFreeVar{ty}\;\HOLSymConst{\HOLTokenIn{}}\;\HOLFreeVar{tys}}} \and
\infer{\HOLinline{(\ensuremath{\HOLFreeVar{ty}\sb{\mathrm{1}}}\HOLSymConst{\ensuremath{\;\to{}}}\ensuremath{\HOLFreeVar{ty}\sb{\mathrm{2}}})\;\HOLSymConst{\HOLTokenIn{}}\;\HOLConst{builtin_closure}\;\HOLFreeVar{tys}}}{\HOLinline{\ensuremath{\HOLFreeVar{ty}\sb{\mathrm{1}}}\;\HOLSymConst{\HOLTokenIn{}}\;\HOLConst{builtin_closure}\;\HOLFreeVar{tys}}&\HOLinline{\ensuremath{\HOLFreeVar{ty}\sb{\mathrm{2}}}\;\HOLSymConst{\HOLTokenIn{}}\;\HOLConst{builtin_closure}\;\HOLFreeVar{tys}}}
\end{mathpar}

\paragraph*{Non-built-ins}
We define operators to collect the non-built-in types of terms and types, and also the non-built-in constants of terms.
The list \HOLinline{\HOLFreeVar{x}\HOLSymConst{\HOLTokenSupBullet{}}} consists of the outermost non-built-in types of a type or term \HOLinline{\HOLFreeVar{x}}.\\[.5em]
\begin{minipage}{.5\textwidth}
  \begin{holthmenv}
    \HOLConst{Bool}\HOLSymConst{\HOLTokenSupBullet{}}\;\HOLTokenDefEquality{}\;\HOLConst{[\,]}\\
(\HOLFreeVar{dom}\HOLSymConst{\ensuremath{\;\to{}}}\HOLFreeVar{rng})\HOLSymConst{\HOLTokenSupBullet{}}\;\HOLTokenDefEquality{}\;\HOLFreeVar{dom}\HOLSymConst{\HOLTokenSupBullet{}}\;\HOLSymConst{++}\;\HOLFreeVar{rng}\HOLSymConst{\HOLTokenSupBullet{}}\\
\HOLFreeVar{ty}\HOLSymConst{\HOLTokenSupBullet{}}\;\HOLTokenDefEquality{}\;[\HOLFreeVar{ty}]\quad\mbox{otherwise}\\[0.5em]
  \end{holthmenv}
\end{minipage}
\begin{minipage}{.5\textwidth}
  \begin{holthmenv}
    (\HOLConst{Var}\;\ensuremath{\HOLFreeVar{v}\sb{\mathrm{0}}}\;\HOLFreeVar{ty})\HOLSymConst{\HOLTokenSupBullet{}}\;\HOLTokenDefEquality{}\;\HOLFreeVar{ty}\HOLSymConst{\HOLTokenSupBullet{}}\\
(\HOLConst{Const}\;\ensuremath{\HOLFreeVar{v}\sb{\mathrm{1}}}\;\HOLFreeVar{ty})\HOLSymConst{\HOLTokenSupBullet{}}\;\HOLTokenDefEquality{}\;\HOLFreeVar{ty}\HOLSymConst{\HOLTokenSupBullet{}}\\
(\HOLConst{Comb}\;\HOLFreeVar{a}\;\HOLFreeVar{b})\HOLSymConst{\HOLTokenSupBullet{}}\;\HOLTokenDefEquality{}\;\HOLFreeVar{a}\HOLSymConst{\HOLTokenSupBullet{}}\;\HOLSymConst{++}\;\HOLFreeVar{b}\HOLSymConst{\HOLTokenSupBullet{}}\\
(\HOLConst{Abs}\;\HOLFreeVar{a}\;\HOLFreeVar{b})\HOLSymConst{\HOLTokenSupBullet{}}\;\HOLTokenDefEquality{}\;\HOLFreeVar{a}\HOLSymConst{\HOLTokenSupBullet{}}\;\HOLSymConst{++}\;\HOLFreeVar{b}\HOLSymConst{\HOLTokenSupBullet{}}
  \end{holthmenv}
\end{minipage}
\vspace{1em}\\
As an example, the outermost non-built-in types of $\mathit{map}_{(\alpha\to\mathsf{Bool})\to\alpha\ \mathsf{list}\to\mathsf{Bool}\ \mathsf{list}}$ over a polymorphic unary list type~$\alpha\ \mathsf{list}$~are:
\begin{holthmenv}
  (\HOLConst{Const}\;\HOLStringLitDG{map}\;((\HOLConst{Tyvar}\;\HOLFreeVar{\ensuremath{\alpha}}\HOLSymConst{\ensuremath{\;\to{}}}\HOLConst{Bool})\HOLSymConst{\ensuremath{\;\to{}}}(\HOLConst{Tyvar}\;\HOLFreeVar{\ensuremath{\alpha}})\HOLSymConst{\;\HOLConst{list}}\HOLSymConst{\ensuremath{\;\to{}}}\HOLConst{Bool}\HOLSymConst{\;\HOLConst{list}}))\HOLSymConst{\HOLTokenSupBullet{}}\;\HOLSymConst{=}\\
\;\;[\HOLConst{Tyvar}\;\HOLFreeVar{\ensuremath{\alpha}};\;(\HOLConst{Tyvar}\;\HOLFreeVar{\ensuremath{\alpha}})\HOLSymConst{\;\HOLConst{list}};\;\HOLConst{Bool}\HOLSymConst{\;\HOLConst{list}}]
\end{holthmenv}
Any type \HOLinline{\HOLFreeVar{ty}} can be recovered from built-in types and the type's outermost non-built-in types:
\begin{center}
  \HOLSymConst{\HOLTokenForall{}}\HOLBoundVar{ty}.\;\HOLBoundVar{ty}\;\HOLSymConst{\HOLTokenIn{}}\;\HOLConst{builtin_closure}\;(\HOLConst{\!\!}\;\HOLBoundVar{ty}\HOLSymConst{\HOLTokenSupBullet{}})
\end{center}
For terms~$t$, we define the list~$t\consts$ to contain all non-built-in constants of~$t$:
\begin{align*}
  &(\HOLConst{Comb}\;\HOLFreeVar{a}\;\HOLFreeVar{b})\HOLSymConst{\HOLTokenSupCirc{}}\;\HOLTokenDefEquality{}\;\HOLFreeVar{a}\HOLSymConst{\HOLTokenSupCirc{}}\;\HOLSymConst{++}\;\HOLFreeVar{b}\HOLSymConst{\HOLTokenSupCirc{}}& &(\HOLConst{Var}\;\HOLFreeVar{x}\;\HOLFreeVar{ty})\HOLSymConst{\HOLTokenSupCirc{}}\;\HOLTokenDefEquality{}\;\HOLConst{[\,]}\\
  &(\HOLConst{Abs}\;\HOLFreeVar{\HOLTokenUnderscore{}}\;\HOLFreeVar{a})\HOLSymConst{\HOLTokenSupCirc{}}\;\HOLTokenDefEquality{}\;\HOLFreeVar{a}\HOLSymConst{\HOLTokenSupCirc{}}& &(\HOLConst{Equal}\;\HOLFreeVar{ty})\HOLSymConst{\HOLTokenSupCirc{}}\;\HOLTokenDefEquality{}\;\HOLConst{[\,]}\\
  &&&(\HOLConst{Const}\;\HOLFreeVar{c}\;\HOLFreeVar{ty})\HOLSymConst{\HOLTokenSupCirc{}}\;\HOLTokenDefEquality{}\;[\HOLConst{Const}\;\HOLFreeVar{c}\;\HOLFreeVar{ty}]\quad\text{otherwise}
\end{align*}

\subsection{Inference system}\label{sec:inference-system}

A \emph{signature} is a pair of functions that assign type constructor names their corresponding arity and constant names their corresponding type.
A \emph{theory} is a pair \HOLinline{(\HOLFreeVar{s}\HOLSymConst{,}\HOLFreeVar{a})} of a signature~\HOLinline{\HOLFreeVar{s}} and a set of terms (axioms)~\HOLinline{\HOLFreeVar{a}}.
Gengelbach and Weber~\cite{DBLP:journals/entcs/GengelbachW18} consider a fixed signature, that is all symbols are initially declared, and a fixed set of axioms.
Here instead, both the signature and the (possibly non-definitional) axioms may be extended (see \cref{sec:theoryextensions}).
The functions $\mathsf{axsof}$, $\mathsf{tysof}$ and $\mathsf{tmsof}$ return the respective components of a theory or signature.

Derivability of \emph{sequents} is defined inductively as a ternary relation~\HOLinline{(\HOLFreeVar{thy}\HOLSymConst{,}\HOLFreeVar{hyps})\;\HOLSymConst{\ensuremath{\vdash}}\;\HOLFreeVar{p}} between a theory~\HOLinline{\HOLFreeVar{thy}}, a list of terms (hypotheses)~\HOLinline{\HOLFreeVar{hyps}} and a term (conclusion)~\HOLinline{\HOLFreeVar{p}}.
We display three of the standard inference rules of higher-order logic, with their syntactic well-formedness constraints.
The condition
\HOLinline{\HOLConst{type_ok}\;(\HOLConst{tysof}\;\HOLFreeVar{ctxt})\;\HOLFreeVar{ty}} requires that \HOLinline{\HOLFreeVar{ty}} is either a type variable or a type constructor applied to the correct number of arguments, as indicated by its arity in the signature, and that these arguments are also \HOLinline{\HOLConst{type_ok}}.
Similarly, \HOLinline{\HOLConst{term_ok}\;(\HOLConst{sigof}\;\HOLFreeVar{thy})\;\HOLFreeVar{p}} requires that \HOLinline{\HOLFreeVar{p}} is a well-typed term, and that its types and constants are instances from the given signature.
Finally, \HOLinline{\HOLConst{theory_ok}\;\HOLFreeVar{ctxt}} requires that in the context \HOLinline{\HOLFreeVar{ctxt}} all axioms are well-formed formulae, the theory has well-typed types and contains at least the built-in symbols.

\begin{mathpar}
\infer[\HOLRuleName{ASSUME}]{\HOLinline{(\HOLFreeVar{thy}\HOLSymConst{,}[\HOLFreeVar{p}])\;\HOLSymConst{\ensuremath{\vdash}}\;\HOLFreeVar{p}}}{\HOLinline{\HOLConst{theory_ok}\;\HOLFreeVar{thy}}&\HOLinline{\HOLFreeVar{p}\;\HOLConst{has_type}\;\HOLConst{Bool}}&\HOLinline{\HOLConst{term_ok}\;(\HOLConst{sigof}\;\HOLFreeVar{thy})\;\HOLFreeVar{p}}}
\and
\infer[\HOLRuleName{ABS}]{\HOLinline{(\HOLFreeVar{thy}\HOLSymConst{,}\HOLConst{[\,]})\;\HOLSymConst{\ensuremath{\vdash}}\;\HOLConst{Comb}\;(\HOLConst{Abs}\;(\HOLConst{Var}\;\HOLFreeVar{x}\;\HOLFreeVar{ty})\;\HOLFreeVar{t})\;(\HOLConst{Var}\;\HOLFreeVar{x}\;\HOLFreeVar{ty})\;\HOLSymConst{===}\;\HOLFreeVar{t}}}{\HOLinline{\HOLConst{theory_ok}\;\HOLFreeVar{thy}}&\HOLinline{\HOLConst{type_ok}\;(\HOLConst{tysof}\;\HOLFreeVar{thy})\;\HOLFreeVar{ty}}&\HOLinline{\HOLConst{term_ok}\;(\HOLConst{sigof}\;\HOLFreeVar{thy})\;\HOLFreeVar{t}}}
\and
\infer[\HOLRuleName{MK_COMB}]{\HOLinline{(\HOLFreeVar{thy}\HOLSymConst{,}\ensuremath{\HOLFreeVar{h}\sb{\mathrm{1}}}\HOLSymConst{\ensuremath{\cup}}\ensuremath{\HOLFreeVar{h}\sb{\mathrm{2}}})\;\HOLSymConst{\ensuremath{\vdash}}\;\HOLConst{Comb}\;\ensuremath{\HOLFreeVar{l}\sb{\mathrm{1}}}\;\ensuremath{\HOLFreeVar{l}\sb{\mathrm{2}}}\;\HOLSymConst{===}\;\HOLConst{Comb}\;\ensuremath{\HOLFreeVar{r}\sb{\mathrm{1}}}\;\ensuremath{\HOLFreeVar{r}\sb{\mathrm{2}}}}}{\HOLinline{(\HOLFreeVar{thy}\HOLSymConst{,}\ensuremath{\HOLFreeVar{h}\sb{\mathrm{1}}})\;\HOLSymConst{\ensuremath{\vdash}}\;\ensuremath{\HOLFreeVar{l}\sb{\mathrm{1}}}\;\HOLSymConst{===}\;\ensuremath{\HOLFreeVar{r}\sb{\mathrm{1}}}}&\HOLinline{(\HOLFreeVar{thy}\HOLSymConst{,}\ensuremath{\HOLFreeVar{h}\sb{\mathrm{2}}})\;\HOLSymConst{\ensuremath{\vdash}}\;\ensuremath{\HOLFreeVar{l}\sb{\mathrm{2}}}\;\HOLSymConst{===}\;\ensuremath{\HOLFreeVar{r}\sb{\mathrm{2}}}}&\HOLinline{\HOLConst{welltyped}\;(\HOLConst{Comb}\;\ensuremath{\HOLFreeVar{l}\sb{\mathrm{1}}}\;\ensuremath{\HOLFreeVar{l}\sb{\mathrm{2}}})}}
\end{mathpar}

\subsection{Theory extensions}\label{sec:theoryextensions}
A theory is obtained from the empty theory by incremental \emph{updates}.
A list of updates is a \emph{context}, and the function $\mathsf{thyof}$ returns the context's theory.
\begin{holthmenv}
  \HOLTyOp{update}\;=\\
\;\;\;\;\HOLConst{NewAxiom}\;\HOLTyOp{term}\\
\;\;\HOLTokenBar{}\;\HOLConst{NewType}\;\HOLTyOp{string}\;\HOLTyOp{num}\\
\;\;\HOLTokenBar{}\;\HOLConst{NewConst}\;\HOLTyOp{string}\;\HOLTyOp{type}\\
\;\;\HOLTokenBar{}\;\HOLConst{TypeDefn}\;\HOLTyOp{string}\;\HOLTyOp{term}\;\HOLTyOp{string}\;\HOLTyOp{string}\\
\;\;\HOLTokenBar{}\;\HOLConst{ConstSpec}\;\HOLTyOp{bool}\;((\HOLTyOp{string}\;\HOLTokenProd{}\;\HOLTyOp{term})\;\HOLTyOp{list})\;\HOLTyOp{term}
\end{holthmenv}
\HOLConst{NewAxiom} adds its argument formula to the theory's set of axioms.
\HOLConst{NewType} and \HOLConst{NewConst} are type and constant \emph{declarations}; they extend the theory's signature.
The remaining \HOLConst{TypeDefn} and \HOLConst{ConstSpec} are \emph{definitions} of a type and of constants, respectively.
Definitions may extend both the signature and the set of axioms, and we defer their discussion to \cref{sec:typedefn,sec:constspec}.

The \HOLinline{\HOLSymConst{updates}} relation specifies when an update is a valid extension of a context:
\begin{holthmenv}
  \infer{\HOLinline{\HOLConst{NewAxiom}\;\HOLFreeVar{prop}\;\HOLConst{updates}\;\HOLFreeVar{ctxt}}}{\begin{array}{c}\HOLinline{\HOLFreeVar{prop}\;\HOLConst{has_type}\;\HOLConst{Bool}}\\\HOLinline{\HOLConst{term_ok}\;(\HOLConst{sigof}\;\HOLFreeVar{ctxt})\;\HOLFreeVar{prop}}\end{array}} \qquad
  \infer{\HOLinline{\HOLConst{NewConst}\;\HOLFreeVar{name}\;\HOLFreeVar{ty}\;\HOLConst{updates}\;\HOLFreeVar{ctxt}}}{\begin{array}{c}\HOLinline{\HOLFreeVar{name}\;\HOLSymConst{\HOLTokenNotIn{}}\;\HOLConst{domain}\;(\HOLConst{tmsof}\;\HOLFreeVar{ctxt})}\\\HOLinline{\HOLConst{type_ok}\;(\HOLConst{tysof}\;\HOLFreeVar{ctxt})\;\HOLFreeVar{ty}}\end{array}} \qquad
  \infer{\HOLinline{\HOLConst{NewType}\;\HOLFreeVar{name}\;\HOLFreeVar{arity}\;\HOLConst{updates}\;\HOLFreeVar{ctxt}}}{\begin{array}{c}\HOLinline{\HOLFreeVar{name}\;\HOLSymConst{\HOLTokenNotIn{}}\;\HOLConst{domain}\;(\HOLConst{tysof}\;\HOLFreeVar{ctxt})}\end{array}}
\end{holthmenv}
The rule for \HOLinline{\HOLConst{NewAxiom}} requires that an axiom is a formula over the context's signature.
The rule for \HOLinline{\HOLConst{NewConst}} requires that the constant's name is new for the context and that its type is from the context's signature.
Similarly, the rule for \HOLinline{\HOLConst{NewType}} requires that the type name is new for the context.

The reflexive relation \HOLinline{\ensuremath{\HOLFreeVar{ctxt}\sb{\mathrm{2}}}\;\HOLConst{extends}\;\ensuremath{\HOLFreeVar{ctxt}\sb{\mathrm{1}}}} expresses that a context \HOLinline{\ensuremath{\HOLFreeVar{ctxt}\sb{\mathrm{2}}}} is obtained from a context \HOLinline{\ensuremath{\HOLFreeVar{ctxt}\sb{\mathrm{1}}}} by a sequence of updates.
The context \HOLinline{\HOLConst{init_ctxt}} contains the built-ins, \ie the types~\HOLinline{\HOLConst{Bool}} and~$\mathsf{Fun}$ and the equality constant.
Its extension \HOLinline{\HOLConst{hol_ctxt}} also contains a type of individuals, the theory of Booleans, a Hilbert-choice constant with its characteristic axiom, and the axioms of extensionality and infinity.

\subsection{Type definitions}\label{sec:typedefn}
\begin{minipage}{0.55\textwidth}
A type definition \HOLinline{\HOLConst{TypeDefn}\;\HOLFreeVar{name}\;\HOLFreeVar{pred}\;\HOLFreeVar{abs}\;\HOLFreeVar{rep}} introduces a new type constructor \HOLinline{\HOLFreeVar{name}} defined by its characteristic, closed predicate \HOLinline{\HOLFreeVar{pred}} as a subset of a host type.
It makes available the type \HOLinline{\HOLConst{Tyapp}\;\HOLFreeVar{name}\;\HOLFreeVar{l}} where the argument list \HOLinline{\HOLFreeVar{l}} corresponds to the distinct type variables of \HOLinline{\HOLFreeVar{pred}}.
A proof that the predicate is satisfiable is a prerequisite, as in HOL types are non-empty.
Additionally, abstraction and representation bijections between the new type and the subset of the host type are axiomatically introduced.
\end{minipage}
\begin{minipage}{0.45\textwidth}
\begin{holthmenv}
  \infer{\HOLinline{\HOLConst{TypeDefn}\;\HOLFreeVar{name}\;\HOLFreeVar{pred}\;\HOLFreeVar{abs}\;\HOLFreeVar{rep}\;\HOLConst{updates}\;\HOLFreeVar{ctxt}}}{\begin{array}{c}\HOLinline{(\HOLConst{thyof}\;\HOLFreeVar{ctxt}\HOLSymConst{,}\HOLConst{[\,]})\;\HOLSymConst{\ensuremath{\vdash}}\;\HOLConst{Comb}\;\HOLFreeVar{pred}\;\HOLFreeVar{witness}}\\\HOLinline{\HOLConst{closed}\;\HOLFreeVar{pred}}\\\HOLinline{\HOLFreeVar{name}\;\HOLSymConst{\HOLTokenNotIn{}}\;\HOLConst{domain}\;(\HOLConst{tysof}\;\HOLFreeVar{ctxt})}\\\HOLinline{\HOLFreeVar{abs}\;\HOLSymConst{\HOLTokenNotIn{}}\;\HOLConst{domain}\;(\HOLConst{tmsof}\;\HOLFreeVar{ctxt})}\\\HOLinline{\HOLFreeVar{rep}\;\HOLSymConst{\HOLTokenNotIn{}}\;\HOLConst{domain}\;(\HOLConst{tmsof}\;\HOLFreeVar{ctxt})}\\\HOLinline{\HOLFreeVar{abs}\;\HOLSymConst{\HOLTokenNotEqual{}}\;\HOLFreeVar{rep}}\end{array}}
\end{holthmenv}
\end{minipage}

\subsection{Constant specification}\label{sec:constspec}

Constant specification \HOLinline{\HOLConst{ConstSpec}\;\HOLFreeVar{ov}\;\HOLFreeVar{eqs}\;\HOLFreeVar{prop}} defines possibly several constants by one axiom \HOLinline{\HOLFreeVar{prop}}.
For~$(c_i,t_i)\in{}eqs$, each of the constants $c_i$ is introduced by a closed witness term $t_i$, that is, the predicate \HOLinline{\HOLFreeVar{prop}} holds assuming all equalities
\begin{align*}
(thy,[\HOLConst{Var}\;c_1\;(\HOLConst{typeof}\;t_1)\;\HOLConst{===}\;t_1;\ldots;\HOLConst{Var}\;c_n\;(\HOLConst{typeof}\;t_n)\;\HOLConst{===}\;t_n])\vdash{}prop\text{.}
\end{align*}
Each of the variables $\HOLConst{Var}\;c_i$ serves as a placeholder for~$\HOLConst{Const}\;c_i$.

If the constant specification is marked as overloading, \ie if \HOLinline{\HOLFreeVar{ov}} is true, the mechanism allows to introduce instances of already declared constants.
Non-overloading constant specifications need to introduce constants with fresh names.
\begin{holthmenv}
  \infer{\HOLinline{\HOLConst{ConstSpec}\;\HOLFreeVar{ov}\;\HOLFreeVar{eqs}\;\HOLFreeVar{prop}\;\HOLConst{updates}\;\HOLFreeVar{ctxt}}}{\begin{array}{c}\HOLinline{(\HOLConst{thyof}\;\HOLFreeVar{ctxt}\HOLSymConst{,}\HOLConst{map}\;(\HOLTokenLambda{}(\HOLBoundVar{s}\HOLSymConst{,}\HOLBoundVar{t}).\;\HOLConst{Var}\;\HOLBoundVar{s}\;(\HOLConst{typeof}\;\HOLBoundVar{t})\;\HOLSymConst{===}\;\HOLBoundVar{t})\;\HOLFreeVar{eqs})\;\HOLSymConst{\ensuremath{\vdash}}\;\HOLFreeVar{prop}}\\\HOLinline{\HOLConst{every}\;(\HOLTokenLambda{}\HOLBoundVar{t}.\;\HOLConst{closed}\;\HOLBoundVar{t}\;\HOLSymConst{\HOLTokenConj{}}\;\HOLSymConst{\HOLTokenForall{}}\HOLBoundVar{v}.\;\HOLBoundVar{v}\;\HOLConst{\HOLTokenIn{}}\;\HOLConst{tvars}\;\HOLBoundVar{t}\;\HOLSymConst{\HOLTokenImp{}}\;\HOLBoundVar{v}\;\HOLConst{\HOLTokenIn{}}\;\HOLConst{tyvars}\;(\HOLConst{typeof}\;\HOLBoundVar{t}))\;(\HOLConst{map}\;\HOLConst{snd}\;\HOLFreeVar{eqs})}\\\HOLinline{\HOLSymConst{\HOLTokenForall{}}\HOLBoundVar{x}\;\HOLBoundVar{ty}.\;\HOLConst{VFREE_IN}\;(\HOLConst{Var}\;\HOLBoundVar{x}\;\HOLBoundVar{ty})\;\HOLFreeVar{prop}\;\HOLSymConst{\HOLTokenImp{}}\;(\HOLBoundVar{x}\HOLSymConst{,}\HOLBoundVar{ty})\HOLConst{\HOLTokenIn{}}\;\HOLConst{map}\;(\HOLTokenLambda{}(\HOLBoundVar{s}\HOLSymConst{,}\HOLBoundVar{t}).\;(\HOLBoundVar{s}\HOLSymConst{,}\HOLConst{typeof}\;\HOLBoundVar{t}))\;\HOLFreeVar{eqs}}\\\HOLinline{\HOLConst{constspec_ok}\;\HOLFreeVar{ov}\;\HOLFreeVar{eqs}\;\HOLFreeVar{prop}\;\HOLFreeVar{ctxt}}\end{array}}
\end{holthmenv}
Here~\HOLinline{\HOLConst{VFREE_IN}\;\HOLFreeVar{x}\;\HOLFreeVar{tm}} denotes that \HOLinline{\HOLFreeVar{x}} is a free term variable in \HOLinline{\HOLFreeVar{tm}}.
The predicate \HOLinline{\HOLConst{constspec_ok}} imposes two important restrictions on constant specifications:
the context resulting from the update needs to be orthogonal (no two defined symbols have a common type instance),
and any introduced overloading of previously declared constants must not allow cycles through the definitions.
We discuss how the latter is avoided with a dependency relation and define orthogonality of contexts in \cref{dependency}.

Constant specification generalises the introduction of new constants via equational axioms, as considered in~\cite{DBLP:journals/entcs/GengelbachW18}, by allowing implicit definitions.\footnote{For instance, Euler's number~$e$ can be implicitly defined as the real-valued solution of a particular differential equation.}
For further discussion of its advantages we refer to~\cite{Art14}.

\subsection{Non-cyclic theories}\label{dependency}
Cycles in theories with overloaded symbols can be avoided by restricting possible definitions in two ways that we define in this section.
First, dependencies introduced by definitions and declarations need to be terminating, which is achieved by Kunčar and Popescu through a dependency relation that Åman Pohjola and Gengelbach~\cite{LPAR23:AmPoGen20} extend to its present form.
Secondly, declared or defined symbols need to be orthogonal~\cite{DBLP:conf/rta/Obua06}, that is any pair of constants or any pair of types that originates from distinct definitions is~orthogonal.

We write $u\equiv{}t$ for definitional updates, to mean that either $u$ is introduced by a type definition with predicate~$t$ or otherwise~$u$ is one of the constants introduced by a constant specification with the witness~$t$.
For a context~\HOLinline{\HOLFreeVar{ctxt}} and types or terms \HOLinline{\HOLFreeVar{u}} and \HOLinline{\HOLFreeVar{v}} the dependency relation \HOLinline{\dependencypair{\;\HOLFreeVar{ctxt}\;}{\;\HOLFreeVar{u}\;}{\;\HOLFreeVar{v}\;}} holds whenever:
\begin{enumerate}
  \item \label{dep_defn}
    There is a definition $u\equiv{}t$ in the context \HOLinline{\HOLFreeVar{ctxt}}
    and
    $v\in{}t\types\cup{}t\consts$, or
  \item \label{dep_cinst} \HOLinline{\HOLFreeVar{u}\;\HOLSymConst{=}\;\HOLConst{Const}\;\HOLFreeVar{\HOLTokenUnderscore{}}\;\HOLFreeVar{ty}} is a constant of type \HOLinline{\HOLFreeVar{ty}} and \HOLinline{\HOLFreeVar{v}\;\HOLConst{\HOLTokenIn{}}\;\HOLFreeVar{ty}\HOLSymConst{\HOLTokenSupBullet{}}}, or
  \item \label{depbitypes} \HOLinline{\HOLFreeVar{u}\;\HOLSymConst{=}\;\HOLConst{Tyapp}\;\HOLFreeVar{\HOLTokenUnderscore{}}\;\HOLFreeVar{l}} is a type
    and \HOLinline{\HOLFreeVar{v}\;\HOLConst{\HOLTokenIn{}}\;\HOLFreeVar{l}}.
\end{enumerate}
The first rule applies only to symbols defined by \HOLinline{\HOLConst{TypeDefn}} or \HOLinline{\HOLConst{ConstSpec}}, whereas the other rules apply also to symbols declared with \HOLinline{\HOLConst{NewType}} and \HOLinline{\HOLConst{NewConst}}.
Formally~$\dep$ is a relation on \HOLinline{\HOLTyOp{type}\;\ensuremath{+}\;\HOLTyOp{term}}, a disjoint union with canonical injections~\HOLinline{\HOLConst{INL}} and~\HOLinline{\HOLConst{INR}}.

The \emph{(type-)substitutive closure} $\mathcal{R}^\downarrow$ of a binary relation \HOLinline{\HOLFreeVar{\ensuremath{\mathcal{R}}}} relates \HOLinline{\HOLConst{\!\!}\;\HOLFreeVar{\varTheta{}}\;\ensuremath{\HOLFreeVar{t}\sb{\mathrm{1}}}} and \HOLinline{\HOLConst{\!\!}\;\HOLFreeVar{\varTheta{}}\;\ensuremath{\HOLFreeVar{t}\sb{\mathrm{2}}}} if $\HOLinline{\ensuremath{\HOLFreeVar{t}\sb{\mathrm{1}}}}\mathrel{\mathcal{R}}\HOLinline{\ensuremath{\HOLFreeVar{t}\sb{\mathrm{2}}}}$.
A relation $\mathcal{R}$ is \emph{terminating} if there is no sequence $(x_i)_{i\in\N}$ such that $x_i\mathrel{\mathcal{R}}x_{i+1}$ for all $i\in\N$.
If a binary relation $\mathcal{R}$ is terminating, its inverse $(\lambda{}x\,y.\,y\mathcal{R}x)$ is well-founded.

A context is \emph{orthogonal} if any two distinct type definitions and any two distinct constant definitions are orthogonal.
Orthogonality ensures that definitional theories have at most one definition for each ground symbol (recall \emph{ground} means type-variable free).

Åman Pohjola and Gengelbach prove that a model exists for each orthogonal context with overloaded definitions whose substitutive closure of the dependency relation is terminating.

\subsection{Semantics}\label{sec:semantics}
In this section we introduce the semantics, which we inherit from Åman Pohjola and Gengelbach~\cite{LPAR23:AmPoGen20}.

\paragraph*{Zermelo-Fraenkel set theory}
The semantics is parametrised on a universe where the axioms of \linebreak Zermelo-Fraenkel set theory (ZF) hold.
A model of ZF is not constructible within HOL by Gödel's incompleteness argument.
This setup is not new~\cite{LPAR23:AmPoGen20}.
The existence of a set-theoretic universe is also an assumption in the mechanised proof of soundness of HOL Light (without overloading)~\cite{KuArMyOw14}, and it originates with Arthan~\cite{spc002}.

Although this parametrisation appears as the assumption \HOLinline{\HOLConst{is_set_theory}\;\HOLFreeVar{mem}} in some theorem statements, in the pretty-printed definitions we often omit the additional argument \HOLinline{\HOLFreeVar{mem}}$\colon$\HOLinline{\ensuremath{\mathcal{U}}\;\ensuremath{\Rightarrow}\;\ensuremath{\mathcal{U}}\;\ensuremath{\Rightarrow}\;\HOLTyOp{bool}}.
Herein, the type variable \HOLinline{\ensuremath{\mathcal{U}}} is the universe of sets.
We also assume \HOLinline{\HOLConst{is_infinite}\;\HOLFreeVar{mem}\;\HOLFreeVar{indset}}, which states that \HOLinline{\HOLFreeVar{indset}}$\colon$\HOLinline{\ensuremath{\mathcal{U}}} is an infinite set.

For set membership \HOLinline{\HOLFreeVar{mem}\;\HOLFreeVar{x}\;\HOLFreeVar{s}} we write \HOLinline{\HOLFreeVar{x}\;\HOLSymConst{\HOLTokenIn{}:}\;\HOLFreeVar{s}}.
\HOLConst{One} is a singleton set, \HOLConst{Boolset} is the set of two distinct elements \HOLinline{\HOLConst{True}} and \HOLinline{\HOLConst{False}}, and \HOLinline{\HOLConst{Boolean}}$\colon$\HOLinline{\HOLTyOp{bool}\;\ensuremath{\Rightarrow}\;\ensuremath{\mathcal{U}}} injects Booleans from HOL into \HOLinline{\ensuremath{\mathcal{U}}} in the expected way.
\HOLinline{\HOLConst{Funspace}\;\HOLFreeVar{s}\;\HOLFreeVar{r}} contains as elements all functions with domain \HOLinline{\HOLFreeVar{s}}$\colon$\HOLinline{\ensuremath{\mathcal{U}}} and co-domain \HOLinline{\HOLFreeVar{r}}$\colon$\HOLinline{\ensuremath{\mathcal{U}}}.
\HOLinline{\HOLConst{Abstract}\;\HOLFreeVar{s}\;\HOLFreeVar{r}\;\HOLFreeVar{f}} is the intersection of the graph of \HOLinline{\HOLFreeVar{f}}$\colon$\HOLinline{\ensuremath{\mathcal{U}}\;\ensuremath{\Rightarrow}\;\ensuremath{\mathcal{U}}} with $s\times{}r$.
In the special case that for any \HOLinline{\HOLFreeVar{x}\;\HOLSymConst{\HOLTokenIn{}:}\;\HOLFreeVar{s}} we have \HOLinline{(\HOLFreeVar{x},\,\HOLFreeVar{f}\;\HOLFreeVar{x})\;\HOLSymConst{\HOLTokenIn{}:}\;\HOLFreeVar{r}}, then \HOLinline{\HOLConst{Abstract}\;\HOLFreeVar{s}\;\HOLFreeVar{r}\;\HOLFreeVar{f}\;\HOLSymConst{\HOLTokenIn{}:}\;\HOLConst{Funspace}\;\HOLFreeVar{s}\;\HOLFreeVar{r}}.
For \HOLinline{\HOLFreeVar{x}\;\HOLSymConst{\HOLTokenIn{}:}\;\HOLFreeVar{s}} and \HOLinline{\HOLFreeVar{g}\;\HOLSymConst{=}\;\HOLConst{Abstract}\;\HOLFreeVar{s}\;\HOLFreeVar{r}\;\HOLFreeVar{f}}, we write \HOLinline{\HOLFreeVar{g}\;\HOLConst{'}\;\HOLFreeVar{x}} for \HOLinline{\HOLFreeVar{f}\;\HOLFreeVar{x}}, namely the second component of \HOLinline{(\HOLFreeVar{x},\,\HOLFreeVar{f}\;\HOLFreeVar{x})} from \HOLinline{\HOLFreeVar{g}}.

\paragraph*{Lazy ground semantics}

A pillar of the semantics is a \emph{(signature) fragment}, which is a tuple \HOLinline{(\HOLFreeVar{tys}\HOLSymConst{,}\HOLFreeVar{consts})} from a signature \HOLinline{\HOLFreeVar{sig}} satisfying:
\begin{holthmenv}
  \HOLConst{is_sig_fragment}\;\HOLFreeVar{sig}\;(\HOLFreeVar{tys}\HOLSymConst{,}\HOLFreeVar{consts})\;\HOLTokenDefEquality{}\\
\;\;\HOLFreeVar{tys}\;\HOLSymConst{\HOLTokenSubset{}}\;\HOLConst{ground_types}\;\HOLFreeVar{sig}\;\HOLSymConst{\HOLTokenConj{}}\;\HOLFreeVar{tys}\;\HOLSymConst{\HOLTokenSubset{}}\;\HOLConst{nonbuiltin_types}\;\HOLSymConst{\HOLTokenConj{}}\;\HOLFreeVar{consts}\;\HOLSymConst{\HOLTokenSubset{}}\;\HOLConst{ground_consts}\;\HOLFreeVar{sig}\;\HOLSymConst{\HOLTokenConj{}}\\
\;\;\HOLFreeVar{consts}\;\HOLSymConst{\HOLTokenSubset{}}\;\HOLConst{nonbuiltin_constinsts}\;\HOLSymConst{\HOLTokenConj{}}\\
\;\;\HOLSymConst{\HOLTokenForall{}}\HOLBoundVar{s}\;\HOLBoundVar{c}.\;(\HOLBoundVar{s}\HOLSymConst{,}\HOLBoundVar{c})\;\HOLSymConst{\HOLTokenIn{}}\;\HOLFreeVar{consts}\;\HOLSymConst{\HOLTokenImp{}}\;\HOLBoundVar{c}\;\HOLSymConst{\HOLTokenIn{}}\;\HOLConst{types_of_frag}\;(\HOLFreeVar{tys}\HOLSymConst{,}\HOLFreeVar{consts})
\end{holthmenv}
The types \HOLinline{\HOLFreeVar{tys}} are ground, non-built-in types from the signature \HOLinline{\HOLFreeVar{sig}}.
Each constant from \HOLinline{\HOLFreeVar{consts}} is non-built-in and has a ground type from the fragment, where \HOLinline{\HOLConst{types_of_frag}\;(\HOLFreeVar{tys}\HOLSymConst{,}\HOLFreeVar{consts})} is defined as the built-in type closure \HOLinline{\HOLConst{builtin_closure}\;\HOLFreeVar{tys}}.
The \emph{total fragment} is the largest fragment of a signature \HOLinline{\HOLFreeVar{sig}}.
\begin{holthmenv}
  \HOLConst{total_fragment}\;\HOLFreeVar{sig}\;\HOLTokenDefEquality{}\;(\HOLConst{ground_types}\;\HOLFreeVar{sig}\;\HOLSymConst{\HOLTokenInter{}}\;\HOLConst{nonbuiltin_types}\HOLSymConst{,}\HOLConst{ground_consts}\;\HOLFreeVar{sig}\;\HOLSymConst{\HOLTokenInter{}}\;\HOLConst{nonbuiltin_constinsts})
\end{holthmenv}

The function \HOLinline{\HOLFreeVar{\ensuremath{\delta}}}$\colon$\HOLinline{\HOLTyOp{type}\;\ensuremath{\Rightarrow}\;\ensuremath{\mathcal{U}}} assigns to each non-built-in type of a fragment a value in the universe.
\HOLinline{\HOLConst{ext}\;\HOLFreeVar{\ensuremath{\delta}}} extends this to built-in types in a standard manner.
Similarly, \HOLinline{\HOLConst{ext}\;\HOLFreeVar{\ensuremath{\gamma}}} extends an interpretation of non-built-in constants \HOLinline{\HOLFreeVar{\ensuremath{\gamma}}} to the built-in constants.
A \emph{(fragment) interpretation} is a tuple \HOLinline{(\HOLFreeVar{\ensuremath{\delta}}\HOLSymConst{,}\HOLFreeVar{\ensuremath{\gamma}})} such that
\begin{holthmenv}
  \HOLConst{is_type_frag_interpretation}\;\HOLFreeVar{tys}\;\HOLFreeVar{\ensuremath{\delta}}\;\HOLTokenDefEquality{}\;\HOLSymConst{\HOLTokenForall{}}\HOLBoundVar{ty}.\;\HOLBoundVar{ty}\;\HOLSymConst{\HOLTokenIn{}}\;\HOLFreeVar{tys}\;\HOLSymConst{\HOLTokenImp{}}\;\HOLConst{inhabited}\;(\HOLFreeVar{\ensuremath{\delta}}\;\HOLBoundVar{ty})\\
  \HOLConst{is_frag_interpretation}\;(\HOLFreeVar{tys}\HOLSymConst{,}\HOLFreeVar{consts})\;\HOLFreeVar{\ensuremath{\delta}}\;\HOLFreeVar{\ensuremath{\gamma}}\;\HOLTokenDefEquality{}\\
\;\;\HOLConst{is_type_frag_interpretation}\;\HOLFreeVar{tys}\;\HOLFreeVar{\ensuremath{\delta}}\;\HOLSymConst{\HOLTokenConj{}}\;\HOLSymConst{\HOLTokenForall{}}(\HOLBoundVar{c}\HOLSymConst{,}\HOLBoundVar{ty}).\;(\HOLBoundVar{c}\HOLSymConst{,}\HOLBoundVar{ty})\;\HOLSymConst{\HOLTokenIn{}}\;\HOLFreeVar{consts}\;\HOLSymConst{\HOLTokenImp{}}\;\HOLFreeVar{\ensuremath{\gamma}}\;(\HOLBoundVar{c}\HOLSymConst{,}\HOLBoundVar{ty})\;\HOLSymConst{\HOLTokenIn{}:}\;\HOLConst{ext}\;\HOLFreeVar{\ensuremath{\delta}}\;\HOLBoundVar{ty}
\end{holthmenv}

\emph{Ground} semantics means that only ground instances of types and constants are interpreted.
A \emph{fragment valuation}~\HOLinline{\HOLFreeVar{v}} assigns to each~\HOLinline{\HOLConst{Var}\;\HOLFreeVar{x}\;\HOLFreeVar{ty}}, with~\HOLinline{\HOLFreeVar{\varTheta{}}\;\HOLFreeVar{ty}} a (ground) type of the fragment, a value that lies in the interpretation of~\HOLinline{\HOLFreeVar{\varTheta{}}\;\HOLFreeVar{ty}}.
\begin{holthmenv}
  \HOLConst{valuates_frag}\;\HOLFreeVar{frag}\;\HOLFreeVar{\ensuremath{\delta}}\;\HOLFreeVar{v}\;\HOLFreeVar{\varTheta{}}\;\HOLTokenDefEquality{}\\
\;\;\HOLSymConst{\HOLTokenForall{}}\HOLBoundVar{x}\;\HOLBoundVar{ty}.\;\HOLConst{\!\!}\;\HOLFreeVar{\varTheta{}}\;\HOLBoundVar{ty}\;\HOLSymConst{\HOLTokenIn{}}\;\HOLConst{types_of_frag}\;\HOLFreeVar{frag}\;\HOLSymConst{\HOLTokenImp{}}\;\HOLFreeVar{v}\;(\HOLBoundVar{x}\HOLSymConst{,}\HOLBoundVar{ty})\;\HOLSymConst{\HOLTokenIn{}:}\;\HOLConst{ext}\;\HOLFreeVar{\ensuremath{\delta}}\;(\HOLConst{\!\!}\;\HOLFreeVar{\varTheta{}}\;\HOLBoundVar{ty})
\end{holthmenv}

The term semantics is defined as a continuation of a fragment interpretation, parametrised by a fragment valuation~\HOLinline{\HOLFreeVar{v}} and a type instantiation~\HOLinline{\HOLFreeVar{\varTheta{}}}.
\begin{holthmenv}
  \HOLConst{termsem}\;\HOLFreeVar{\ensuremath{\delta}}\;\HOLFreeVar{\ensuremath{\gamma}}\;\HOLFreeVar{v}\;\HOLFreeVar{\varTheta{}}\;(\HOLConst{Var}\;\HOLFreeVar{x}\;\HOLFreeVar{ty})\;\HOLTokenDefEquality{}\;\HOLFreeVar{v}\;(\HOLFreeVar{x}\HOLSymConst{,}\HOLFreeVar{ty})\\
  \HOLConst{termsem}\;\HOLFreeVar{\ensuremath{\delta}}\;\HOLFreeVar{\ensuremath{\gamma}}\;\HOLFreeVar{v}\;\HOLFreeVar{\varTheta{}}\;(\HOLConst{Const}\;\HOLFreeVar{name}\;\HOLFreeVar{ty})\;\HOLTokenDefEquality{}\;\HOLFreeVar{\ensuremath{\gamma}}\;(\HOLFreeVar{name}\HOLSymConst{,}\HOLConst{\!\!}\;\HOLFreeVar{\varTheta{}}\;\HOLFreeVar{ty})\\
  \HOLConst{termsem}\;\HOLFreeVar{\ensuremath{\delta}}\;\HOLFreeVar{\ensuremath{\gamma}}\;\HOLFreeVar{v}\;\HOLFreeVar{\varTheta{}}\;(\HOLConst{Comb}\;\ensuremath{\HOLFreeVar{t}\sb{\mathrm{1}}}\;\ensuremath{\HOLFreeVar{t}\sb{\mathrm{2}}})\;\HOLTokenDefEquality{}\;\HOLConst{termsem}\;\HOLFreeVar{\ensuremath{\delta}}\;\HOLFreeVar{\ensuremath{\gamma}}\;\HOLFreeVar{v}\;\HOLFreeVar{\varTheta{}}\;\ensuremath{\HOLFreeVar{t}\sb{\mathrm{1}}}\;\HOLConst{'}\;(\HOLConst{termsem}\;\HOLFreeVar{\ensuremath{\delta}}\;\HOLFreeVar{\ensuremath{\gamma}}\;\HOLFreeVar{v}\;\HOLFreeVar{\varTheta{}}\;\ensuremath{\HOLFreeVar{t}\sb{\mathrm{2}}})\\
  \HOLConst{termsem}\;\HOLFreeVar{\ensuremath{\delta}}\;\HOLFreeVar{\ensuremath{\gamma}}\;\HOLFreeVar{v}\;\HOLFreeVar{\varTheta{}}\;(\HOLConst{Abs}\;(\HOLConst{Var}\;\HOLFreeVar{x}\;\HOLFreeVar{ty})\;\HOLFreeVar{b})\;\HOLTokenDefEquality{}\\
\;\;\HOLConst{Abstract}\;(\HOLFreeVar{\ensuremath{\delta}}\;(\HOLConst{\!\!}\;\HOLFreeVar{\varTheta{}}\;\HOLFreeVar{ty}))\;(\HOLFreeVar{\ensuremath{\delta}}\;(\HOLConst{\!\!}\;\HOLFreeVar{\varTheta{}}\;(\HOLConst{typeof}\;\HOLFreeVar{b})))\;(\HOLTokenLambda{}\HOLBoundVar{m}.\;\HOLConst{termsem}\;\HOLFreeVar{\ensuremath{\delta}}\;\HOLFreeVar{\ensuremath{\gamma}}\;\HOLFreeVar{v}\ensuremath{\llparenthesis}(\HOLFreeVar{x}\HOLSymConst{,}\HOLFreeVar{ty})\;\mapsto\;\HOLBoundVar{m}\ensuremath{\rrparenthesis}\;\HOLFreeVar{\varTheta{}}\;\HOLFreeVar{b})
\end{holthmenv}
Herein \HOLinline{\HOLFreeVar{f}\ensuremath{\llparenthesis}\HOLFreeVar{x}\;\mapsto\;\HOLFreeVar{y}\ensuremath{\rrparenthesis}} is the function that at \HOLinline{\HOLFreeVar{x}} takes the value \HOLinline{\HOLFreeVar{y}} and elsewhere equals \HOLinline{\HOLFreeVar{f}}.

The semantics applies type substitutions \emph{lazily}, \ie as late as possible and never to the type of term variables.
This avoids a problem~\cite{LPAR23:AmPoGen20} with the eager semantics of Kunčar and Popescu: in HOL's Church-style atoms, variables~\HOLinline{\HOLConst{Var}\;\HOLFreeVar{x}\;(\HOLConst{Tyvar}\;\HOLFreeVar{a})} and~\HOLinline{\HOLConst{Var}\;\HOLFreeVar{x}\;\HOLConst{Bool}} are distinct and hence should be allowed to have different valuations under all type substitutions.
With the lazy ground semantics, for \HOLinline{\HOLFreeVar{\varTheta{}}\;(\HOLConst{Tyvar}\;\HOLFreeVar{a})\;\HOLSymConst{=}\;\HOLConst{Bool}} we just have~\HOLinline{\HOLFreeVar{v}\;(\HOLFreeVar{x}\HOLSymConst{,}\HOLConst{Tyvar}\;\HOLFreeVar{a})\;\HOLSymConst{\HOLTokenIn{}:}\;\HOLConst{ext}\;\HOLFreeVar{\ensuremath{\delta}}\;(\HOLFreeVar{\varTheta{}}\;(\HOLConst{Tyvar}\;\HOLFreeVar{a}))}$=\HOLConst{Boolset}$ and \HOLinline{\HOLFreeVar{v}\;(\HOLFreeVar{x}\HOLSymConst{,}\HOLConst{Bool})\;\HOLSymConst{\HOLTokenIn{}:}\;\HOLConst{ext}\;\HOLFreeVar{\ensuremath{\delta}}\;(\HOLFreeVar{\varTheta{}}\;\HOLConst{Bool})}$=\HOLConst{Boolset}$.
In contrast, eager ground semantics erroneously identifies \HOLinline{\HOLFreeVar{v}\;(\HOLFreeVar{x}\HOLSymConst{,}\HOLFreeVar{\varTheta{}}\;(\HOLConst{Tyvar}\;\HOLFreeVar{a}))\;\HOLSymConst{=}\;\HOLFreeVar{v}\;(\HOLFreeVar{x}\HOLSymConst{,}\HOLFreeVar{\varTheta{}}\;\HOLConst{Bool})}.

We define the satisfaction relation of a fragment interpretation \HOLinline{(\HOLFreeVar{\ensuremath{\delta}}\HOLSymConst{,}\HOLFreeVar{\ensuremath{\gamma}})}, hypotheses \HOLinline{\HOLFreeVar{hyps}} and a term~\HOLinline{\HOLFreeVar{p}} \wrt a fragment \HOLinline{\HOLFreeVar{frag}} and a type substitution~\HOLinline{\HOLFreeVar{\varTheta{}}}.
Every fragment valuation $v$ that satisfies all instantiated hypotheses must satisfy the instantiated term \HOLinline{\HOLFreeVar{\varTheta{}}\;\HOLFreeVar{p}}.
\begin{holthmenv}
  \HOLConst{satisfies}\;\HOLFreeVar{frag}\;\HOLFreeVar{\ensuremath{\delta}}\;\HOLFreeVar{\ensuremath{\gamma}}\;\HOLFreeVar{\varTheta{}}\;(\HOLFreeVar{hyps}\HOLSymConst{,}\HOLFreeVar{p})\;\HOLTokenDefEquality{}\\
\;\;\HOLSymConst{\HOLTokenForall{}}\HOLBoundVar{v}.\\
\;\;\;\;\;\;\HOLConst{valuates_frag}\;\HOLFreeVar{frag}\;\HOLFreeVar{\ensuremath{\delta}}\;\HOLBoundVar{v}\;\HOLFreeVar{\varTheta{}}\;\HOLSymConst{\HOLTokenConj{}}\;\HOLFreeVar{p}\;\HOLSymConst{\HOLTokenIn{}}\;\HOLConst{terms_of_frag_uninst}\;\HOLFreeVar{frag}\;\HOLFreeVar{\varTheta{}}\;\HOLSymConst{\HOLTokenConj{}}\\
\;\;\;\;\;\;\HOLConst{every}\;(\HOLTokenLambda{}\HOLBoundVar{t}.\;\HOLBoundVar{t}\;\HOLSymConst{\HOLTokenIn{}}\;\HOLConst{terms_of_frag_uninst}\;\HOLFreeVar{frag}\;\HOLFreeVar{\varTheta{}})\;\HOLFreeVar{hyps}\;\HOLSymConst{\HOLTokenConj{}}\;\HOLConst{every}\;(\HOLTokenLambda{}\HOLBoundVar{t}.\;\HOLConst{termsem}\;\HOLFreeVar{\ensuremath{\delta}}\;\HOLFreeVar{\ensuremath{\gamma}}\;\HOLBoundVar{v}\;\HOLFreeVar{\varTheta{}}\;\HOLBoundVar{t}\;\HOLSymConst{=}\;\HOLConst{True})\;\HOLFreeVar{hyps}\;\HOLSymConst{\HOLTokenImp{}}\\
\;\;\;\;\;\;\;\;\HOLConst{termsem}\;\HOLFreeVar{\ensuremath{\delta}}\;\HOLFreeVar{\ensuremath{\gamma}}\;\HOLBoundVar{v}\;\HOLFreeVar{\varTheta{}}\;\HOLFreeVar{p}\;\HOLSymConst{=}\;\HOLConst{True}
\end{holthmenv}
Satisfaction of hypotheses \HOLinline{\HOLFreeVar{hyps}} and a conclusion \HOLinline{\HOLFreeVar{p}} \wrt a fragment interpretation \HOLinline{(\HOLFreeVar{\ensuremath{\delta}}\HOLSymConst{,}\HOLFreeVar{\ensuremath{\gamma}})} and a signature \HOLinline{\HOLFreeVar{sig}} is quantified over all ground type substitutions of the signature.
\begin{holthmenv}
  \HOLConst{sat}\;\HOLFreeVar{sig}\;\HOLFreeVar{\ensuremath{\delta}}\;\HOLFreeVar{\ensuremath{\gamma}}\;(\HOLFreeVar{hyps}\HOLSymConst{,}\HOLFreeVar{p})\;\HOLTokenDefEquality{}\\
\;\;\HOLSymConst{\HOLTokenForall{}}\HOLBoundVar{\varTheta{}}.\\
\;\;\;\;\;\;(\HOLSymConst{\HOLTokenForall{}}\HOLBoundVar{ty}.\;\HOLConst{tyvars}\;(\HOLBoundVar{\varTheta{}}\;\HOLBoundVar{ty})\;\HOLSymConst{=}\;\HOLConst{[\,]})\;\HOLSymConst{\HOLTokenConj{}}\;(\HOLSymConst{\HOLTokenForall{}}\HOLBoundVar{ty}.\;\HOLConst{type_ok}\;(\HOLConst{tysof}\;\HOLFreeVar{sig})\;(\HOLBoundVar{\varTheta{}}\;\HOLBoundVar{ty}))\;\HOLSymConst{\HOLTokenConj{}}\\
\;\;\;\;\;\;\HOLConst{every}\;(\HOLTokenLambda{}\HOLBoundVar{tm}.\;\HOLBoundVar{tm}\;\HOLSymConst{\HOLTokenIn{}}\;\HOLConst{ground_terms_uninst}\;\HOLFreeVar{sig}\;\HOLBoundVar{\varTheta{}})\;\HOLFreeVar{hyps}\;\HOLSymConst{\HOLTokenConj{}}\;\HOLFreeVar{p}\;\HOLSymConst{\HOLTokenIn{}}\;\HOLConst{ground_terms_uninst}\;\HOLFreeVar{sig}\;\HOLBoundVar{\varTheta{}}\;\HOLSymConst{\HOLTokenImp{}}\\
\;\;\;\;\;\;\;\;\HOLConst{satisfies}\;(\HOLConst{total_fragment}\;\HOLFreeVar{sig})\;\HOLFreeVar{\ensuremath{\delta}}\;\HOLFreeVar{\ensuremath{\gamma}}\;\HOLBoundVar{\varTheta{}}\;(\HOLFreeVar{hyps}\HOLSymConst{,}\HOLFreeVar{p})
\end{holthmenv}
A total fragment interpretation \HOLinline{(\HOLFreeVar{\ensuremath{\delta}}\HOLSymConst{,}\HOLFreeVar{\ensuremath{\gamma}})} is a model of a theory \HOLinline{\HOLFreeVar{thy}} if all of the theory's axioms are satisfied.
\begin{holthmenv}
  \HOLConst{models}\;\HOLFreeVar{\ensuremath{\delta}}\;\HOLFreeVar{\ensuremath{\gamma}}\;\HOLFreeVar{thy}\;\HOLTokenDefEquality{}\\
\;\;\HOLConst{is_frag_interpretation}\;(\HOLConst{total_fragment}\;(\HOLConst{sigof}\;\HOLFreeVar{thy}))\;\HOLFreeVar{\ensuremath{\delta}}\;\HOLFreeVar{\ensuremath{\gamma}}\;\HOLSymConst{\HOLTokenConj{}}\\
\;\;\HOLSymConst{\HOLTokenForall{}}\HOLBoundVar{p}.\;\HOLBoundVar{p}\;\HOLSymConst{\HOLTokenIn{}}\;\HOLConst{axsof}\;\HOLFreeVar{thy}\;\HOLSymConst{\HOLTokenImp{}}\;\HOLConst{sat}\;(\HOLConst{sigof}\;\HOLFreeVar{thy})\;(\HOLConst{ext}\;\HOLFreeVar{\ensuremath{\delta}})\;(\HOLConst{ext}\;(\HOLConst{ext}\;\HOLFreeVar{\ensuremath{\delta}})\;\HOLFreeVar{\ensuremath{\gamma}})\;(\HOLConst{[\,]}\HOLSymConst{,}\HOLBoundVar{p})
\end{holthmenv}
As the semantic counterpart of derivability (\cref{sec:inference-system}), we define semantic entailment \HOLinline{(\HOLFreeVar{thy}\HOLSymConst{,}\HOLFreeVar{hyps})\;\HOLSymConst{\ensuremath{\vDash}}\;\HOLFreeVar{p}}.
\begin{holthmenv}
  (\HOLFreeVar{thy}\HOLSymConst{,}\HOLFreeVar{hyps})\;\HOLSymConst{\ensuremath{\vDash}}\;\HOLFreeVar{p}\;\HOLTokenDefEquality{}\\
\;\;\HOLConst{theory_ok}\;\HOLFreeVar{thy}\;\HOLSymConst{\HOLTokenConj{}}\;\HOLConst{every}\;(\HOLConst{term_ok}\;(\HOLConst{sigof}\;\HOLFreeVar{thy}))\;(\HOLFreeVar{p}\HOLSymConst{::}\HOLFreeVar{hyps})\;\HOLSymConst{\HOLTokenConj{}}\\
\;\;\HOLConst{every}\;(\HOLTokenLambda{}\HOLBoundVar{p}.\;\HOLBoundVar{p}\;\HOLConst{has_type}\;\HOLConst{Bool})\;(\HOLFreeVar{p}\HOLSymConst{::}\HOLFreeVar{hyps})\;\HOLSymConst{\HOLTokenConj{}}\;\HOLConst{hypset_ok}\;\HOLFreeVar{hyps}\;\HOLSymConst{\HOLTokenConj{}}\\
\;\;\HOLSymConst{\HOLTokenForall{}}\HOLBoundVar{\ensuremath{\delta}}\;\HOLBoundVar{\ensuremath{\gamma}}.\;\HOLConst{models}\;\HOLBoundVar{\ensuremath{\delta}}\;\HOLBoundVar{\ensuremath{\gamma}}\;\HOLFreeVar{thy}\;\HOLSymConst{\HOLTokenImp{}}\;\HOLConst{sat}\;(\HOLConst{sigof}\;\HOLFreeVar{thy})\;(\HOLConst{ext}\;\HOLBoundVar{\ensuremath{\delta}})\;(\HOLConst{ext}\;(\HOLConst{ext}\;\HOLBoundVar{\ensuremath{\delta}})\;\HOLBoundVar{\ensuremath{\gamma}})\;(\HOLFreeVar{hyps}\HOLSymConst{,}\HOLFreeVar{p})
\end{holthmenv}
The inference system is sound \wrt this semantics~\cite{LPAR23:AmPoGen20}.

\section{Symbol-independent fragment}\label{sec:independentfragment}

After recapitulating the syntax and semantics in the previous section, we are set to discuss our contribution.
The convenience that constants may be used prior to their definition comes at the price that interpretations of previously introduced symbols may change in extensions that define previously undefined symbols.
For instance, the interpretation may change for defined orderings on lists, lexicographically defined as $\leq_{\alpha\;\mathsf{list}\to\alpha\;\mathsf{list}\to\bool}$ \HOLConst{===} $\mathsf{lex}(\leq_{\alpha\to\alpha\to\bool})$, if an update defines any previously undefined instance of $\leq$.
In this section we carve out the fragment of all symbols that are unaffected by a theory update.

An \emph{independent fragment} collects constants and types of a host fragment \HOLinline{\HOLFreeVar{frag}} whose definitions within a theory context \HOLinline{\HOLFreeVar{ctxt}} are independent of any of the symbols from a set $U$.
\begin{holthmenv}
  \HOLConst{indep_frag}\;\HOLFreeVar{ctxt}\;\HOLFreeVar{U}\;\HOLFreeVar{frag}\;\HOLTokenDefEquality{}\\
\;\;\HOLKeyword{let}\;\HOLBoundVar{V}\;=\;\HOLTokenLeftbrace{}\HOLBoundVar{x}\;\HOLTokenBar{}\;\HOLSymConst{\HOLTokenExists{}}\HOLBoundVar{\varTheta{}}\;\HOLBoundVar{u}.\;\HOLBoundVar{u}\;\HOLConst{\HOLTokenIn{}}\;\HOLFreeVar{U}\;\HOLSymConst{\HOLTokenConj{}}\;\dependencyrtcpair{\;\HOLFreeVar{ctxt}\;}{\;\HOLBoundVar{x}\;}{\;\HOLConst{\!\!}\;\HOLBoundVar{\varTheta{}}\;\HOLBoundVar{u}\;}\HOLTokenRightbrace{}\;;\\
\;\;\;\;\;\;\;\;\ensuremath{\HOLBoundVar{V}\sb{\mathrm{2}}}\;=\;\HOLTokenLeftbrace{}(\HOLBoundVar{x}\HOLSymConst{,}\HOLBoundVar{ty})\;\HOLTokenBar{}\;\HOLConst{INR}\;(\HOLConst{Const}\;\HOLBoundVar{x}\;\HOLBoundVar{ty})\;\HOLSymConst{\HOLTokenIn{}}\;\HOLBoundVar{V}\HOLTokenRightbrace{}\;;\;\ensuremath{\HOLBoundVar{V}\sb{\mathrm{1}}}\;=\;\HOLTokenLeftbrace{}\HOLBoundVar{x}\;\HOLTokenBar{}\;\HOLConst{INL}\;\HOLBoundVar{x}\;\HOLSymConst{\HOLTokenIn{}}\;\HOLBoundVar{V}\HOLTokenRightbrace{}\;\HOLKeyword{in}\\
\;\;\;\;(\HOLConst{fst}\;\HOLFreeVar{frag}\;\HOLConst{\ensuremath{\setminus{}}}\;\ensuremath{\HOLBoundVar{V}\sb{\mathrm{1}}}\HOLSymConst{,}\HOLConst{snd}\;\HOLFreeVar{frag}\;\HOLConst{\ensuremath{\setminus{}}}\;\ensuremath{\HOLBoundVar{V}\sb{\mathrm{2}}})\\[0.5em]
\end{holthmenv}
The set~$U$ contains the symbols introduced by a theory extension.
In contrast to~\cite{DBLP:journals/entcs/GengelbachW18}, where~$U$ is a singleton set, we allow the introduction of several symbols at once, \eg via constant specification.
The set $V$ is the pre-image of type instances~\HOLinline{\HOLFreeVar{\varTheta{}}\;\HOLFreeVar{u}} of elements~$u$ from~$U$ (with \HOLinline{\HOLFreeVar{\varTheta{}}} a ground type substitution) under the reflexive-transitive, type-substitutive closure of the dependency relation~$\dep_{ctxt}$.
As host fragment \HOLinline{\HOLFreeVar{frag}}, we only consider total fragments (over different signatures).
An independent fragment of a total fragment is indeed a signature fragment, since constants depend on their types.
\begin{holthmenv}
  \HOLTokenTurnstile{}\;\HOLFreeVar{ctxt}\;\HOLConst{extends}\;\HOLConst{init_ctxt}\;\HOLSymConst{\HOLTokenImp{}}\\
\;\;\;\;\;\HOLConst{is_sig_fragment}\;(\HOLConst{sigof}\;\HOLFreeVar{ctxt})\;(\HOLConst{indep_frag}\;\HOLFreeVar{ctxt}\;\HOLFreeVar{U}\;(\HOLConst{total_fragment}\;(\HOLConst{sigof}\;\HOLFreeVar{ctxt})))
\end{holthmenv}

We prove this claim in script, to give a flavour of the reasoning involved in the mechanisation.
Thereby we amend the earlier proof~\cite{DBLP:journals/entcs/GengelbachW18} for the case where a type substitution $\rho$ and $\types$ do not
commute on a type $\varsigma$, \ie $\rho(\varsigma\types)\neq\rho(\varsigma)\types$.
(This case had been excluded by a faulty lemma inherited from Kunčar and Popescu.)

For a fixed context, $F_U$ denotes the fragment independent of symbols $U$, and $\GType\types$ and $\GCInst\consts$ are all types and non-built-in constants of the total fragment, respectively.

\begin{proof}
  For a ground constant instance $c_\sigma\in\GCInst\consts\setminus{}V$,
  we show that also its type $\sigma$ is from the types of~$F_U$.
  Assume that~$\sigma\notin\HOLConst{builtin_closure}(\GType\types\setminus{}V)$.
  Thus $\sigma\types\not\subseteq\GType\types\setminus{}V$
  and there is a type~$\tau\in\sigma\types\cap{}V$. Assuming the dependency
  $c_\sigma\dep^{\downarrow+}\tau$ the contradiction $c_\sigma\in{}V$ follows.
  We now show $c_\sigma\dep^{\downarrow+}\tau$ for $\tau\in\sigma\types$: \\
  Let $c_\varsigma$ be a (defined or declared) constant. It holds
  $c_\varsigma\dep{}t$ for $t\in\varsigma\types$ and thus for any instance
  $c_{\rho(\varsigma)}\dep^\downarrow\rho(t)$ for $t\in\varsigma\types$.
  Generally, $\rho(\varsigma)\types\neq\rho(\varsigma\types)$
  as Åman Pohjola and Gengelbach notice~\cite{LPAR23:AmPoGen20}.
  If $\varsigma$ is a type variable or a non-built-in type,
  $\varsigma\types=\{\varsigma\}$, then
  $c_{\rho(\varsigma)}\dep^\downarrow\rho(\varsigma)$ and
  $\rho(\varsigma)\dep{}t$ for $t\in\rho(\varsigma)\types$.
  If on the other hand $\varsigma=a\to{}b$ is the built-in function type,
  thus $\sigma$ is a function type and let $\rho$ be such that
  $\rho(a\to{}b)=\sigma$. Any type below $\sigma$ and above
  $\tau\in\sigma\types$ is a function type (as
  $\tau\in\sigma\types\neq\{\sigma\}$).
  If $\tau$ is introduced by a type instantiation, then within $a\to{}b$ there is a type variable $\alpha$ such that $\rho(\alpha)$ syntactically contains $\tau$.
  Thus $c_{a\to{}b}\dep\alpha$ by $\alpha\in(a\to{}b)\types$ and $\rho(\alpha)\dep^+\tau$ (as in $\rho(\alpha)$ there are only function types above $\tau$).
  If $\tau$ was not introduced by a type instantiation and $\tau'$ is the type within $a\to{}b$ such that $\rho(\tau')=\tau$, then $c_{a\to{}b}\dep\tau'$ and consequently $c_{\rho(a\to{}b)}\dep^\downarrow\rho(\tau')=\tau$.
\end{proof}

\paragraph*{Symbols introduced by a theory extension}
Until now, the independent fragment has been defined without regard to
the theory extension mechanism, to contain all symbols that are
independent of the symbols from an arbitrary set $U$.
The relevant independent fragments are those that are independent of a theory extension, \ie those for which $U$ contains the constant instances and types that are introduced by a theory update.
For an update $\HOLinline{\HOLFreeVar{upd}}$, we set $U=\HOLinline{\HOLConst{upd_introduces}\;\HOLFreeVar{upd}}$ as the apex of the independent fragment cone.
\begin{holthmenv}
\HOLConst{upd_introduces}\;(\HOLConst{ConstSpec}\;\HOLFreeVar{ov}\;\HOLFreeVar{eqs}\;\HOLFreeVar{prop})\;\HOLTokenDefEquality{}\;\HOLConst{map}\;(\HOLTokenLambda{}(\HOLBoundVar{s}\HOLSymConst{,}\HOLBoundVar{t}).\;\HOLConst{INR}\;(\HOLConst{Const}\;\HOLBoundVar{s}\;(\HOLConst{typeof}\;\HOLBoundVar{t})))\;\HOLFreeVar{eqs}\\
\HOLConst{upd_introduces}\;(\HOLConst{TypeDefn}\;\HOLFreeVar{name}\;\HOLFreeVar{pred}\;\HOLFreeVar{abs}\;\HOLFreeVar{rep})\;\HOLTokenDefEquality{}\\
\;\;[\HOLConst{INL}\;(\HOLConst{Tyapp}\;\HOLFreeVar{name}\;(\HOLConst{map}\;\HOLConst{Tyvar}\;(\HOLConst{mlstring_sort}\;(\HOLConst{tvars}\;\HOLFreeVar{pred}))))]\\
\HOLConst{upd_introduces}\;(\HOLConst{NewType}\;\HOLFreeVar{name}\;\HOLFreeVar{arity})\;\HOLTokenDefEquality{}\\
\;\;[\HOLConst{INL}\;(\HOLConst{Tyapp}\;\HOLFreeVar{name}\;(\HOLConst{map}\;\HOLConst{Tyvar}\;(\HOLConst{genlist}\;(\HOLTokenLambda{}\HOLBoundVar{x}.\;\HOLConst{implode}\;(\HOLConst{replicate}\;(\HOLConst{SUC}\;\HOLBoundVar{x})\;\HOLCharLit{a}))\;\HOLFreeVar{arity})))]\\
\HOLConst{upd_introduces}\;(\HOLConst{NewConst}\;\HOLFreeVar{name}\;\HOLFreeVar{ty})\;\HOLTokenDefEquality{}\;[\HOLConst{INR}\;(\HOLConst{Const}\;\HOLFreeVar{name}\;\HOLFreeVar{ty})]\\
\HOLConst{upd_introduces}\;(\HOLConst{NewAxiom}\;\HOLFreeVar{prop})\;\HOLTokenDefEquality{}\;\HOLConst{[\,]}
\end{holthmenv}
For constant specifications and declarations, \HOLinline{\HOLConst{upd_introduces}} returns the constants available for use after the theory update.
For type definitions, the introduced type constructor has as arguments all type variables of the defining predicate sorted by name.
Type declarations introduce a type constructor whose arguments are \emph{arity} many distinct type variables.

In the definition of \HOLinline{\HOLConst{upd_introduces}} we make two choices:
\begin{itemize}
  \item
    The independent fragment of an update defining a type $\tau$ by a predicate $t_{\sigma\to\bool}$ defines $U=\{\tau\}$.
    For a type substitution $\rho$ either all instances $\rho(\tau)$, $\rep_{\rho(\sigma\to\tau)}$ and $\abs_{\rho(\tau\to\sigma)}$ are in $F_U$ or otherwise in its complement (that we earlier denoted $V$).
    Although the proof of said property is non-trivial, the choice of defining $U=\{\tau\}$ instead of $U=\{\tau,\rep_{\sigma\to\tau},\abs_{\tau\to\sigma}\}$ adds the convenience (for case analysis in some proofs) that any constant introduced by an update (\wrt \HOLinline{\HOLConst{upd_introduces}}) does not come from a type definition.
  \item
    As non-definitional axioms generally are not conservative, any symbol's
    interpretation may be affected by such an update, hence we define
     \HOLConst{upd_introduces}\;(\HOLConst{NewAxiom}\;\HOLFreeVar{prop})\;\HOLTokenDefEquality{}\;\HOLConst{[\,]}.
\end{itemize}

\noindent
We henceforth only regard independent fragments related to theory updates.
\begin{holthmenv}
  \HOLConst{indep_frag_upd}\;\HOLFreeVar{ctxt}\;\HOLFreeVar{upd}\;\HOLFreeVar{frag}\;\HOLTokenDefEquality{}\;\HOLConst{indep_frag}\;\HOLFreeVar{ctxt}\;(\HOLConst{upd_introduces}\;\HOLFreeVar{upd})\;\HOLFreeVar{frag}\\[0.5em]
\end{holthmenv}
The independent fragment of a theory \HOLinline{\HOLFreeVar{ctxt}} extended by \HOLinline{\HOLFreeVar{upd}} is carved out from the total fragment over the extended signature, but factually any symbols introduced by the update are not within the fragment:
\begin{holthmenv}
  \HOLTokenTurnstile{}\;\HOLKeyword{let}\;\HOLBoundVar{idf}\;=\;\HOLConst{indep_frag_upd}\;(\HOLFreeVar{upd}\HOLSymConst{::}\HOLFreeVar{ctxt})\;\HOLFreeVar{upd}\;(\HOLConst{total_fragment}\;(\HOLConst{sigof}\;(\HOLFreeVar{upd}\HOLSymConst{::}\HOLFreeVar{ctxt})))\;\HOLKeyword{in}\\
\;\;\;\;\;\HOLFreeVar{upd}\HOLSymConst{::}\HOLFreeVar{ctxt}\;\HOLConst{extends}\;\HOLConst{init_ctxt}\;\HOLSymConst{\HOLTokenImp{}}\\
\;\;\;\;\;\;\;\HOLConst{fst}\;\HOLBoundVar{idf}\;\HOLSymConst{\HOLTokenSubset{}}\;\HOLConst{fst}\;(\HOLConst{total_fragment}\;(\HOLConst{sigof}\;\HOLFreeVar{ctxt}))\;\HOLSymConst{\HOLTokenConj{}}\;\HOLConst{snd}\;\HOLBoundVar{idf}\;\HOLSymConst{\HOLTokenSubset{}}\;\HOLConst{snd}\;(\HOLConst{total_fragment}\;(\HOLConst{sigof}\;\HOLFreeVar{ctxt}))
\end{holthmenv}
Hereby, the \HOLinline{\HOLFreeVar{upd}}-independent fragments over the signatures \HOLinline{\HOLFreeVar{ctxt}} and \HOLinline{\HOLFreeVar{upd}\HOLSymConst{::}\HOLFreeVar{ctxt}} are equal, as  each symbol introduced by the extension by~\HOLinline{\HOLFreeVar{upd}} depends on a symbol in \HOLinline{\HOLConst{upd_introduces}\;\HOLFreeVar{upd}}.

\section{Model-theoretic Conservativity}\label{sec:conservativity}

In this section we discuss how we construct a model of an extended theory while keeping parts of a model from the theory prior to extension.
With the properties of the construction and an extra assumption on the given models we prove model-theoretic conservativity.

\subsection{Model construction}\label{modelconstruction}
From a model $(\varDelta,\varGamma)$ of a theory $\HOLinline{\HOLFreeVar{ctxt}}$ we construct a model $(\delta,\gamma)$ of the extension $\HOLinline{\HOLFreeVar{upd}\HOLSymConst{::}\HOLFreeVar{ctxt}}$.
Supported theory extensions are either extensions by definition or declaration of constants or a type, or otherwise admissible non-definitional axioms.
A model of the extended theory is constructed by recursion over part of the~$\dep^\downarrow$ relation, based on the model~$(\Delta,\Gamma)$.
In contrast, the model construction in~\cite{DBLP:journals/jar/KuncarP19a} obtains a model from the ground up, by recursion over the entire~$\dep^\downarrow$ relation, without reference to any previous~interpretation.

A model is constructed by two mutually recursive functions \HOLinline{\HOLConst{type_interpretation_ext}\;\HOLFreeVar{ind}\;\HOLFreeVar{ctxt}\;\HOLFreeVar{upd}\;\HOLFreeVar{\varDelta{}}\;\HOLFreeVar{\varGamma{}}\;\HOLFreeVar{ty}} and \HOLinline{\HOLConst{term_interpretation_ext}\;\HOLFreeVar{ind}\;\HOLFreeVar{ctxt}\;\HOLFreeVar{upd}\;\HOLFreeVar{\varDelta{}}\;\HOLFreeVar{\varGamma{}}\;\HOLFreeVar{c}\;\HOLFreeVar{ty}} that return the interpretation of a type or constant instance, respectively.
As arguments these functions take the model $(\varDelta,\varGamma)$ of the theory~\HOLinline{\HOLFreeVar{ctxt}}, the update~$\HOLinline{\HOLFreeVar{upd}}$ and an infinite type~$\HOLinline{\HOLFreeVar{ind}}$.
The model construction for a definitional theory extension $\HOLinline{\HOLFreeVar{upd}\HOLSymConst{::}\HOLFreeVar{ctxt}}$ is guarded with a check:
if the symbol to interpret lies in the independent fragment
$$\HOLinline{\HOLConst{indep_frag_upd}\;(\HOLFreeVar{upd}\HOLSymConst{::}\HOLFreeVar{ctxt})\;\HOLFreeVar{upd}\;(\HOLConst{total_fragment}\;(\HOLConst{sigof}\;\HOLFreeVar{ctxt}))}$$
of a definitional update $\HOLinline{\HOLFreeVar{upd}}$, the symbol may be interpreted \wrt the model $(\varDelta,\varGamma)$.
Otherwise the symbol's interpretation is constructed as discussed in earlier work~\cite{LPAR23:AmPoGen20,DBLP:journals/jar/KuncarP19a,DBLP:journals/entcs/GengelbachW18}.

Our amendments to the model construction are a few lines each (here the four lines of the second $\mathsf{if}$~branch) in \HOLinline{\HOLConst{type_interpretation_ext}} and \HOLinline{\HOLConst{term_interpretation_ext}}.
The inherited tedious parts are~elided.
\begin{holthmenv}
  \HOLConst{type_interpretation_ext}\;\HOLFreeVar{ind}\;\HOLFreeVar{upd}\;\HOLFreeVar{ctxt}\;\HOLFreeVar{\varDelta{}}\;\HOLFreeVar{\varGamma{}}\;\HOLFreeVar{ty}\;\HOLTokenDefEquality{}\\
\;\;\HOLKeyword{if}\;\;\HOLSymConst{\HOLTokenNeg{}}\HOLConst{wellformed}\;(\HOLFreeVar{upd}\HOLSymConst{::}\HOLFreeVar{ctxt})\\
\;\;\HOLKeyword{then}\;\;\HOLConst{One}\\
\;\;\HOLKeyword{else}\;\HOLKeyword{if}\\
\;\;\;\;(\HOLSymConst{\HOLTokenForall{}}\HOLBoundVar{tm}.\;\HOLFreeVar{upd}\;\HOLSymConst{\HOLTokenNotEqual{}}\;\HOLConst{NewAxiom}\;\HOLBoundVar{tm})\;\HOLSymConst{\HOLTokenConj{}}\\
\;\;\;\;\HOLFreeVar{ty}\;\HOLSymConst{\HOLTokenIn{}}\;\HOLConst{fst}\;(\HOLConst{indep_frag_upd}\;(\HOLFreeVar{upd}\HOLSymConst{::}\HOLFreeVar{ctxt})\;\HOLFreeVar{upd}\;(\HOLConst{total_fragment}\;(\HOLConst{sigof}\;\HOLFreeVar{ctxt})))\\
\;\;\HOLKeyword{then}\;\;\HOLFreeVar{\varDelta{}}\;\HOLFreeVar{ty}\\
\;\;\HOLKeyword{else}\;\HOLConst{\ensuremath{\dots}}
\end{holthmenv}

\paragraph*{Requirements for Constant Specification}
The differing constant definition mechanism entails that the model construction yields no model, but only a fragment interpretation of the theory's total fragment.

For a theory \HOLinline{\HOLFreeVar{ctxt}} that has a model $(\Delta,\Gamma)$ we need to prove that any axiom from \HOLinline{\HOLFreeVar{ctxt}} holds in a model~$(\delta,\gamma)$ of a valid theory extension \HOLinline{\HOLFreeVar{upd}\HOLSymConst{::}\HOLFreeVar{ctxt}}.
In its proof we are presented with a sub-case that occurs due to the different definitional mechanism for constants, as compared to Gengelbach and Weber.
We illustrate the problem by an example theory:

Let \HOLinline{\HOLFreeVar{ctxt}} be a theory where by constant specification two constants $d_\bool$ and $e_\bool$ are defined to be distinct by the axiom $d_\bool\neq{}e_\bool$, that holds for the witnesses $d_\bool=\mathsf{False}$ and $e_\bool=(c_\bool\Rightarrow\mathsf{True})$ for a declared-only constant $c_\bool$.
Let an update \HOLinline{\HOLFreeVar{upd}} define $c_\bool=\mathsf{True}$.
For the fragment of \HOLinline{\HOLFreeVar{ctxt}} that is independent of this update of $c_\bool$ we write~$F$.
For a model $(\varDelta,\varGamma)$ of \HOLinline{\HOLFreeVar{ctxt}} we have to show that the axiom $d_\bool\neq{}e_\bool$ holds in the model extension $(\delta,\gamma)$ for \HOLinline{\HOLFreeVar{upd}\HOLSymConst{::}\HOLFreeVar{ctxt}} as obtained from the above model construction.
With $d_\bool\in{}F$ and $e_\bool\not\in{}F$, it is impossible to prove that $\gamma(d_\bool)\neq\gamma(e_\bool)$ as we only know $\gamma(d_\bool)=\varGamma(d_\bool)$ and $\gamma(e_\bool)=\mathsf{true}$.

In the example two constants are simultaneously introduced and defined in terms of another, and only one lies in the independent fragment.
In the iterative model construction, information is lost on how constants are interpreted that are dependencies of a symbol.
We choose to only extend models where each defined constant is interpreted as its witness.

We require that all constants, defined by constant specifications in a context \HOLinline{\HOLFreeVar{ctxt}}, are interpreted equal to their witness in a model $(\varDelta,\varGamma)$.
\begin{holthmenv}
  \HOLConst{models_witnesses}\;\HOLFreeVar{\varDelta{}}\;\HOLFreeVar{\varGamma{}}\;\HOLFreeVar{ctxt}\;\HOLTokenDefEquality{}\\
\;\;\HOLSymConst{\HOLTokenForall{}}\HOLBoundVar{ov}\;\HOLBoundVar{cl}\;\HOLBoundVar{prop}\;\HOLBoundVar{c}\;\HOLBoundVar{cdefn}\;\HOLBoundVar{ty}\;\HOLBoundVar{\varTheta{}}.\\
\;\;\;\;\;\;\HOLConst{ConstSpec}\;\HOLBoundVar{ov}\;\HOLBoundVar{cl}\;\HOLBoundVar{prop}\;\HOLConst{\HOLTokenIn{}}\;\HOLFreeVar{ctxt}\;\HOLSymConst{\HOLTokenConj{}}\;(\HOLBoundVar{c}\HOLSymConst{,}\HOLBoundVar{cdefn})\HOLConst{\HOLTokenIn{}}\;\HOLBoundVar{cl}\;\HOLSymConst{\HOLTokenConj{}}\;\HOLBoundVar{ty}\;\HOLSymConst{=}\;\HOLConst{typeof}\;\HOLBoundVar{cdefn}\;\HOLSymConst{\HOLTokenConj{}}\\
\;\;\;\;\;\;(\HOLBoundVar{c}\HOLSymConst{,}\HOLConst{\!\!}\;\HOLBoundVar{\varTheta{}}\;\HOLBoundVar{ty})\;\HOLSymConst{\HOLTokenIn{}}\;\HOLConst{ground_consts}\;(\HOLConst{sigof}\;\HOLFreeVar{ctxt})\;\HOLSymConst{\HOLTokenConj{}}\;(\HOLBoundVar{c}\HOLSymConst{,}\HOLConst{\!\!}\;\HOLBoundVar{\varTheta{}}\;\HOLBoundVar{ty})\;\HOLSymConst{\HOLTokenIn{}}\;\HOLConst{nonbuiltin_constinsts}\;\HOLSymConst{\HOLTokenImp{}}\\
\;\;\;\;\;\;\;\;\HOLFreeVar{\varGamma{}}\;(\HOLBoundVar{c}\HOLSymConst{,}\HOLConst{\!\!}\;\HOLBoundVar{\varTheta{}}\;\HOLBoundVar{ty})\;\HOLSymConst{=}\;\HOLConst{termsem}\;(\HOLConst{ext}\;\HOLFreeVar{\varDelta{}})\;(\HOLConst{ext}\;(\HOLConst{ext}\;\HOLFreeVar{\varDelta{}})\;\HOLFreeVar{\varGamma{}})\;\HOLConst{empty_valuation}\;\HOLBoundVar{\varTheta{}}\;\HOLBoundVar{cdefn}
\end{holthmenv}
This added requirement is preserved by the model construction.
\begin{holthmenv}
  \HOLTokenTurnstile{}\;\HOLConst{is_set_theory}\;\HOLFreeVar{mem}\;\HOLSymConst{\HOLTokenImp{}}\\
\;\;\;\;\;\HOLSymConst{\HOLTokenForall{}}\HOLBoundVar{upd}\;\HOLBoundVar{ctxt}\;\HOLBoundVar{\varDelta{}}\;\HOLBoundVar{\varGamma{}}.\\
\;\;\;\;\;\;\;\;\;\HOLBoundVar{upd}\HOLSymConst{::}\HOLBoundVar{ctxt}\;\HOLConst{extends}\;\HOLConst{init_ctxt}\;\HOLSymConst{\HOLTokenConj{}}\;\HOLConst{inhabited}\;\HOLFreeVar{ind}\;\HOLSymConst{\HOLTokenConj{}}\\
\;\;\;\;\;\;\;\;\;\HOLConst{is_frag_interpretation}\;(\HOLConst{total_fragment}\;(\HOLConst{sigof}\;\HOLBoundVar{ctxt}))\;\HOLBoundVar{\varDelta{}}\;\HOLBoundVar{\varGamma{}}\;\HOLSymConst{\HOLTokenConj{}}\\
\;\;\;\;\;\;\;\;\;\HOLConst{models_witnesses}\;\HOLBoundVar{\varDelta{}}\;\HOLBoundVar{\varGamma{}}\;\HOLBoundVar{ctxt}\;\HOLSymConst{\HOLTokenImp{}}\\
\;\;\;\;\;\;\;\;\;\;\;\HOLConst{models_witnesses}\;(\HOLConst{type_interpretation_ext}\;\HOLFreeVar{ind}\;\HOLBoundVar{upd}\;\HOLBoundVar{ctxt}\;\HOLBoundVar{\varDelta{}}\;\HOLBoundVar{\varGamma{}})\\
\;\;\;\;\;\;\;\;\;\;\;\;\;(\HOLConst{term_interpretation_ext}\;\HOLFreeVar{ind}\;\HOLBoundVar{upd}\;\HOLBoundVar{ctxt}\;\HOLBoundVar{\varDelta{}}\;\HOLBoundVar{\varGamma{}})\;(\HOLBoundVar{upd}\HOLSymConst{::}\HOLBoundVar{ctxt})
\end{holthmenv}

The restriction to models of theories that satisfy \HOLinline{\HOLConst{models_witnesses}} keeps the expressivity of Arthan's constant specification and is conservative \wrt constant definition.

Alternatively, the problem as depicted in the example can be circumvented through extending the dependency relation with \emph{cross-dependencies}.
For simultaneously introduced constants $d$ and $e$, any dependency $x$ of $e$ (\ie $e\dep{}x$) also becomes a dependency of $d$ (\ie $d\dep{}x$) and likewise with $d$ and $e$ swapped.
Any constants that are introduced together would thereby be assumed to be related, which reduces expressivity.
For two declared constants $f_\alpha$ and $g_\bool$ the otherwise legitimate simultaneous definition of $f_\alpha={}g_\alpha$ and $g_\bool=\mathsf{True}$ becomes impossible, as it is cyclic:
$g_\bool\dep^\downarrow{}g_\bool$.
Instead each conjunct would need to be a theory extension on its own.

\subsection{Model-theoretic conservativity}

In this subsection we introduce our main result.
The mechanism to extend theories by definitions or declarations is model-theoretically conservative if for any theory \HOLinline{\HOLFreeVar{ctxt}} with a model $(\varDelta,\varGamma)$ that interprets any constant witness pair from constant specification equal, any theory extension \HOLinline{\HOLFreeVar{upd}\HOLSymConst{::}\HOLFreeVar{ctxt}} (where \HOLinline{\HOLFreeVar{upd}} is a definition or declaration) has a model $(\delta,\gamma)$ that also satisfies the property:
\begin{align*}
  \HOLKeyword{let}\;\HOLBoundVar{idf}\;=\;\HOLConst{indep_frag_upd}\;(\HOLFreeVar{upd}\HOLSymConst{::}\HOLFreeVar{ctxt})\;\HOLFreeVar{upd}\;(\HOLConst{total_fragment}\;(\HOLConst{sigof}\;\HOLFreeVar{ctxt}))\;\HOLKeyword{in}\\
\;\;(\HOLSymConst{\HOLTokenForall{}}\HOLBoundVar{ty}.\;\HOLBoundVar{ty}\;\HOLSymConst{\HOLTokenIn{}}\;\HOLConst{fst}\;\HOLBoundVar{idf}\;\HOLSymConst{\HOLTokenImp{}}\;\HOLFreeVar{\ensuremath{\delta}}\;\HOLBoundVar{ty}\;\HOLSymConst{=}\;\HOLFreeVar{\varDelta{}}\;\HOLBoundVar{ty})\;\HOLSymConst{\HOLTokenConj{}}\;\HOLSymConst{\HOLTokenForall{}}\HOLBoundVar{c}\;\HOLBoundVar{ty}.\;(\HOLBoundVar{c}\HOLSymConst{,}\HOLBoundVar{ty})\;\HOLSymConst{\HOLTokenIn{}}\;\HOLConst{snd}\;\HOLBoundVar{idf}\;\HOLSymConst{\HOLTokenImp{}}\;\HOLFreeVar{\ensuremath{\gamma}}\;(\HOLBoundVar{c}\HOLSymConst{,}\HOLBoundVar{ty})\;\HOLSymConst{=}\;\HOLFreeVar{\varGamma{}}\;(\HOLBoundVar{c}\HOLSymConst{,}\HOLBoundVar{ty})
\end{align*}
If this property holds, it naturally extends to any ground term that is built from built-in types and symbols from the independent fragment \HOLinline{\HOLFreeVar{idf}}. Hence any such term's interpretation in the new model~$(\delta,\gamma)$ equals its interpretation in the old model $(\varDelta,\varGamma)$.
With the restriction to models that interpret constants as their witnesses we derive that the construction in \cref{modelconstruction} yields a model.
\begin{holthmenv}
  \HOLTokenTurnstile{}\;\HOLConst{is_set_theory}\;\HOLFreeVar{mem}\;\HOLSymConst{\HOLTokenImp{}}\\
\;\;\;\;\;\HOLSymConst{\HOLTokenForall{}}\HOLBoundVar{upd}\;\HOLBoundVar{ctxt}\;\HOLBoundVar{\varDelta{}}\;\HOLBoundVar{\varGamma{}}.\\
\;\;\;\;\;\;\;\;\;\HOLBoundVar{ctxt}\;\HOLConst{extends}\;\HOLConst{init_ctxt}\;\HOLSymConst{\HOLTokenConj{}}\;\HOLConst{inhabited}\;\HOLFreeVar{ind}\;\HOLSymConst{\HOLTokenConj{}}\;\HOLBoundVar{upd}\;\HOLConst{updates}\;\HOLBoundVar{ctxt}\;\HOLSymConst{\HOLTokenConj{}}\\
\;\;\;\;\;\;\;\;\;\HOLConst{axioms_admissible}\;\HOLFreeVar{mem}\;\HOLFreeVar{ind}\;(\HOLBoundVar{upd}\HOLSymConst{::}\HOLBoundVar{ctxt})\;\HOLSymConst{\HOLTokenConj{}}\;\HOLConst{models}\;\HOLBoundVar{\varDelta{}}\;\HOLBoundVar{\varGamma{}}\;(\HOLConst{thyof}\;\HOLBoundVar{ctxt})\;\HOLSymConst{\HOLTokenConj{}}\;\HOLConst{models_witnesses}\;\HOLBoundVar{\varDelta{}}\;\HOLBoundVar{\varGamma{}}\;\HOLBoundVar{ctxt}\;\HOLSymConst{\HOLTokenImp{}}\\
\;\;\;\;\;\;\;\;\;\;\;\HOLConst{models}\;(\HOLConst{type_interpretation_ext}\;\HOLFreeVar{ind}\;\HOLBoundVar{upd}\;\HOLBoundVar{ctxt}\;\HOLBoundVar{\varDelta{}}\;\HOLBoundVar{\varGamma{}})\;(\HOLConst{term_interpretation_ext}\;\HOLFreeVar{ind}\;\HOLBoundVar{upd}\;\HOLBoundVar{ctxt}\;\HOLBoundVar{\varDelta{}}\;\HOLBoundVar{\varGamma{}})\\
\;\;\;\;\;\;\;\;\;\;\;\;\;(\HOLConst{thyof}\;(\HOLBoundVar{upd}\HOLSymConst{::}\HOLBoundVar{ctxt}))
\end{holthmenv}
This constructed model trivially satisfies the given property that interpretations from the \HOLinline{\HOLFreeVar{upd}}-independent fragment are kept if the \HOLinline{\HOLFreeVar{upd}} is a declaration or a definition.

At different stages in the proof of model-theoretic conservativity, case analysis occurs of how an update~\HOLinline{\HOLFreeVar{upd}} may extend a theory \HOLinline{\HOLFreeVar{ctxt}} by \HOLinline{\HOLFreeVar{upd}\;\HOLConst{updates}\;\HOLFreeVar{ctxt}}.
As an example, proof obligations similar to the following reoccur frequently in the formalisation.
To show that a symbol \HOLinline{\HOLFreeVar{x}} keeps its interpretation in a model extension \wrt an update~\HOLinline{\HOLFreeVar{upd}}, one has to show that \HOLinline{\HOLFreeVar{x}} is independent of the update~\HOLinline{\HOLFreeVar{upd}} by proving that all dependencies of \HOLinline{\HOLFreeVar{x}} are on symbols from the \HOLinline{\HOLFreeVar{upd}}-independent~fragment.

Future work could investigate if the model construction may be conservative even \wrt \HOLinline{\HOLConst{NewAxiom}} updates of admissible axioms from \HOLinline{\HOLConst{hol_ctxt}}.

\subsection{Consistency}

As a consequence of the model construction from the previous section, we obtain consistency of \emph{definitional} extensions of \HOLinline{\HOLConst{hol_ctxt}}, that is extensions that do not contain \HOLinline{\HOLConst{NewAxiom}}.

A theory is \emph{consistent} if there is a provable and an unprovable sequent.
We inherit the following definition from Kumar \etal~\cite{DBLP:journals/jar/KumarAMO16}.
\begin{holthmenv}
  \HOLConst{consistent_theory}\;\HOLFreeVar{thy}\;\HOLTokenDefEquality{}\\
\;\;(\HOLFreeVar{thy}\HOLSymConst{,}\HOLConst{[\,]})\;\HOLSymConst{\ensuremath{\vdash}}\;\HOLConst{Var}\;\HOLStringLitDG{x}\;\HOLConst{Bool}\;\HOLSymConst{===}\;\HOLConst{Var}\;\HOLStringLitDG{x}\;\HOLConst{Bool}\;\HOLSymConst{\HOLTokenConj{}}\;\HOLSymConst{\HOLTokenNeg{}}((\HOLFreeVar{thy}\HOLSymConst{,}\HOLConst{[\,]})\;\HOLSymConst{\ensuremath{\vdash}}\;\HOLConst{Var}\;\HOLStringLitDG{x}\;\HOLConst{Bool}\;\HOLSymConst{===}\;\HOLConst{Var}\;\HOLStringLitDG{y}\;\HOLConst{Bool})
\end{holthmenv}
As a corollary of our work, the existence of a model of \HOLinline{\HOLConst{init_ctxt}} combined with the incremental model construction yields consistency of definitional extensions of \HOLinline{\HOLConst{hol_ctxt}}~\cite{DBLP:journals/jar/KuncarP19a,LPAR23:AmPoGen20}.
The restriction on the interpretations of constants as their witnesses trivially holds in \HOLinline{\HOLConst{init_ctxt}} and is an invariant in the induction.
\begin{holthmenv}
  \HOLTokenTurnstile{}\;\HOLConst{is_set_theory}\;\HOLFreeVar{mem}\;\HOLSymConst{\HOLTokenConj{}}\;\HOLConst{is_infinite}\;\HOLFreeVar{mem}\;\HOLFreeVar{ind}\;\HOLSymConst{\HOLTokenImp{}}\\
\;\;\;\;\;\HOLSymConst{\HOLTokenForall{}}\HOLBoundVar{ctxt}.\;\HOLConst{definitional_extension}\;\HOLBoundVar{ctxt}\;\HOLConst{hol_ctxt}\;\HOLSymConst{\HOLTokenImp{}}\;\HOLConst{consistent_theory}\;(\HOLConst{thyof}\;\HOLBoundVar{ctxt})
\end{holthmenv}
This work thus generalises and replaces the earlier non-incremental model construction~\cite{LPAR23:AmPoGen20}.

\section{Related Work}\label{sec:relatedwork}

For untyped first-order logic, extension by definition of predicate and function symbols is discussed by Shoenfield~\cite[\S~4.6]{Sho67}.
A definitions by a predicate extends a theory with an equivalence that contains the predicate only on the left-hand side; a definition by a function symbol requires the proof that the function symbol indeed is a mathematical function.
These mechanisms are proof-theoretically conservative, and each model of the original theory has one unique corresponding model of the extended theory.
In consequence, both definitional mechanisms are model-theoretically conservative.

Farmer~\cite{farmer94} defines an extension of a theory to be a super-set that is a model-theoretic conservative extension, hence keeps model interpretation and consistency.
By example of simply-typed first-order logic with extension by algebraic datatypes and constant definitions, the author discusses also weaker notions of semantic conservativity and its properties \wrt theory embeddings, so called theory instantiation.

In their formalisation of HOL Light without overloading~\cite{DBLP:journals/jar/KumarAMO16}, Kumar \etal also make model-theoretic conservativity a requirement for theory extension by definitions or declarations.
They denote this property \HOLConst{sound_update}\;\HOLFreeVar{ctxt}\;\HOLFreeVar{upd} of each such extension of \HOLinline{\HOLFreeVar{ctxt}} by \HOLinline{\HOLFreeVar{upd}}, and prove consistency by an inductive argument.
As the definition mechanism for constants they use constant specification that allows to introduce multiple constants at once, given witnesses for which the defining axiom is derivable.
Constant specification was first introduced by Arthan~\cite{Art14}, and is is implemented in HOL4~\cite{HOL14} and ProofPower.

The study of theoretical foundations of overloaded definitions (together with type classes in higher-order logic) dates back to Wenzel~\cite{DBLP:conf/tphol/Wenzel97}.
For Wenzel an extension mechanism for deductive logics needs to be syntactically conservative, which he proves for constant definition where all instances are defined at once.
In addition, the considered constant definitions can be unfolded, which is called \emph{realisability}.
In this discussion the interplay of overloaded constants and type definitions is not considered.

To avoid inconsistencies Obua~\cite{DBLP:conf/rta/Obua06} remarks that the unfolding of definitions needs to terminate for both type and constant definitions.
Further Obua discusses that termination is not semi-decidable for overloaded definitions that recurse through types.
The proof sketch of conservativity of overloading in Isabelle, he misses that inconsistencies may be introduced by dependencies through types.

For the Isabelle framework with its Haskell-style type classes, Wenzel and Haftmann~\cite{DBLP:conf/types/HaftmannW06} state requirements on overloading definitions without discussing if these suffice for acyclic dependencies.

Kunčar and Popescu~\cite{DBLP:journals/jar/KuncarP19a} aim to close the consistency gap for definitional theories in Isabelle, in showing that every definitional theory has a model, by a model construction that recurses into the dependencies of definitions.
Fixable gaps in their result are closed in the mechanisation of the model construction by Åman Pohjola and Gengelbach~\cite{LPAR23:AmPoGen20}.
Instead of constant definition their mechanisation considers Arthan's constant specification, and gives the above discussed \emph{lazy fragment-ground} semantics.
We base on their implementation work and generalise their monolithic model construction.

In two works, Kunčar and Popescu study consistency of definitional theories by syntactic arguments.
They encode formulas through an unfolding of definitions into a richer logic HOL with comprehension types~(HOLC) and prove that provability is preserved~\cite{KuncarPopescu2017a}.
Ultimately, definitional theories are consistent by the consistency of~HOLC.

In another paper, they use an unfolding that stays in the logic of HOL~\cite{DBLP:journals/pacmpl/Kuncar018} by relativising defined types in formulas to a predicate on the defined type's host type.
The proof-theoretic conservativity result holds for any definitional theory unfolded into initial HOL, and motivates a dual model-theoretic conservativity result where any model of initial HOL can be extended to a model of a given definitional theory.
Our paper proves model-theoretic conservativity of two arbitrary definitional theories.

In recent work Gengelbach and Weber~\cite{GengelbachW20} prove model-theoretic conservativity of definitional theories~\cite{DBLP:journals/entcs/GengelbachW18} for semantics that do not require full function spaces in order to derive their syntactic counterparts.
A definitional extension of a theory is proof-theoretically conservative, that is, if a formula's types and constants are unchanged by a theory update, and the formula is derivable after the update, then it is also derivable from the theory before the update.
Their proof-theoretical result holds for constant definition and it is unclear how that result is transferable to constant specification with regard to the additional restriction on models \HOLinline{\HOLConst{models_witnesses}} in our proof.

Mizar is a theorem prover that supports overloading of symbols even for overlapping sub-types~\cite{DBLP:journals/jfrea/GrabowskiKN10}, where either the interpretation \wrt a definition may be specified or the most recently introduced definition is chosen for interpretation.
Despite mentions of consistency of this sophisticated mechanism (\eg~\cite{Rudnicki92}) there is no proof for consistency or conservativity of Mizar.

\section{Conclusion}\label{sec:conclusion}

We established that type definitions and constant specifications in HOL with ad-hoc overloading of arbitrary theories above \HOLinline{\HOLConst{init_ctxt}} with fixed admissible axioms from \HOLinline{\HOLConst{hol_ctxt}} are model-theoretically conservative.
The result holds for models that interpret each constant introduced by constant specification equal to the constant's witness.
An interpretation of this result is that the definitional mechanisms of Isabelle/HOL are semantically speaking robustly designed: at least symbols that are independent of an update may keep their interpretation in a model extension.

Model-theoretic conservativity has a proof-theoretic (syntactic) counterpart.
Roughly, an extension is \emph{proof-theoretically conservative} if it entails no new theorems in the original language.
In other words, every formula of the original language that is a theorem in the extension is already provable in the original~theory.

In earlier work, Kunčar and Popescu~\cite{DBLP:journals/pacmpl/Kuncar018} show that any definitional theory is a proof-theoretically conservative extension of \emph{initial HOL}, \ie \HOLinline{\HOLConst{hol_ctxt}}.
The semantic counterpart is that any definitional theory is model-theoretically conservative above initial HOL.
Comparably, our semantic conservativity is stronger as it holds for arbitrary theories above \HOLinline{\HOLConst{hol_ctxt}}, which we achieved by utilising the independent fragment, \ie a subset of the signature that is independent of a theory extension.

We conjecture that the syntactic counterpart of our result holds: if~$D'$ is an extension of~$D$ such that $D'\vdash\phi$, where~$\phi$ is a formula whose non-built-in constant instances and types are independent of symbols defined in $D'\setminus D$, then $D\vdash\phi$.  Gengelbach and Weber recently proved this conjecture for constant definition through equality axioms~\cite{GengelbachW20}.  We leave its study for the more general constant specification mechanism~\cite{Art14} to future work.


\bibliographystyle{eptcs}
\bibliography{paper}

\begin{thebibliography}{10}
\providecommand{\bibitemdeclare}[2]{}
\providecommand{\surnamestart}{}
\providecommand{\surnameend}{}
\providecommand{\urlprefix}{Available at }
\providecommand{\url}[1]{\texttt{#1}}
\providecommand{\href}[2]{\texttt{#2}}
\providecommand{\urlalt}[2]{\href{#1}{#2}}
\providecommand{\doi}[1]{doi:\urlalt{http://dx.doi.org/#1}{#1}}
\providecommand{\bibinfo}[2]{#2}

\bibitemdeclare{unpublished}{spc002}
\bibitem{spc002}
\bibinfo{author}{Rob \surnamestart Arthan\surnameend}:
  \emph{\bibinfo{title}{{HOL} Formalised: Semantics}}.
\newblock \urlprefix\url{http://www.lemma-one.com/ProofPower/specs/spc002.pdf}.

\bibitemdeclare{inproceedings}{Art14}
\bibitem{Art14}
\bibinfo{author}{Rob \surnamestart Arthan\surnameend} (\bibinfo{year}{2014}):
  \emph{\bibinfo{title}{{{HOL Constant Definition Done Right}}}}.
\newblock In: {\sl \bibinfo{booktitle}{Interactive {{Theorem Proving}}}},
  \bibinfo{publisher}{{Springer International Publishing}}, pp.
  \bibinfo{pages}{531--536}, \doi{10.1007/978-3-319-08970-6_34}.

\bibitemdeclare{unpublished}{farmer94}
\bibitem{farmer94}
\bibinfo{author}{William~M. \surnamestart Farmer\surnameend}:
  \emph{\bibinfo{title}{A General Method for Safely Overwriting Theories in
  Mechanized Mathematics Systems}}.
\newblock \urlprefix\url{http://imps.mcmaster.ca/doc/overwriting-theories.pdf}.

\bibitemdeclare{inproceedings}{DBLP:journals/entcs/GengelbachW18}
\bibitem{DBLP:journals/entcs/GengelbachW18}
\bibinfo{author}{Arve \surnamestart Gengelbach\surnameend} \&
  \bibinfo{author}{Tjark \surnamestart Weber\surnameend}
  (\bibinfo{year}{2017}): \emph{\bibinfo{title}{Model-Theoretic {C}onservative
  {E}xtension for {D}efinitional {T}heories}}.
\newblock In \bibinfo{editor}{Sandra \surnamestart Alves\surnameend} \&
  \bibinfo{editor}{Renata \surnamestart Wasserman\surnameend}, editors: {\sl
  \bibinfo{booktitle}{12th Workshop on Logical and Semantic Frameworks, with
  Applications, {LSFA} 2017, Bras{\'{\i}}lia, Brazil, September 23-24, 2017}},
  {\sl \bibinfo{series}{Electronic Notes in Theoretical Computer Science}}
  \bibinfo{volume}{338}, \bibinfo{publisher}{Elsevier}, pp.
  \bibinfo{pages}{133--145}, \doi{10.1016/j.entcs.2018.10.009}.

\bibitemdeclare{inproceedings}{GengelbachW20}
\bibitem{GengelbachW20}
\bibinfo{author}{Arve \surnamestart Gengelbach\surnameend} \&
  \bibinfo{author}{Tjark \surnamestart Weber\surnameend}
  (\bibinfo{year}{2020}): \emph{\bibinfo{title}{{Proof-theoretic Conservativity
  for HOL with Ad-hoc Overloading}}}.
\newblock In \bibinfo{editor}{Violet Ka~I \surnamestart Pun\surnameend},
  \bibinfo{editor}{Volker \surnamestart Stolz\surnameend} \&
  \bibinfo{editor}{Adenilso \surnamestart da~Silva~Sim{\~{a}}o\surnameend},
  editors: {\sl \bibinfo{booktitle}{Theoretical Aspects of Computing - {ICTAC}
  2020 - 17th International Colloquium, Macau, China, November 30 - December 4,
  2020, Proceedings}}, {\sl \bibinfo{series}{Lecture Notes in Computer
  Science}} \bibinfo{volume}{12545}, \bibinfo{publisher}{Springer}, pp.
  \bibinfo{pages}{23--42}, \doi{10.1007/978-3-030-64276-1_2}.

\bibitemdeclare{article}{DBLP:journals/jfrea/GrabowskiKN10}
\bibitem{DBLP:journals/jfrea/GrabowskiKN10}
\bibinfo{author}{Adam \surnamestart Grabowski\surnameend},
  \bibinfo{author}{Artur \surnamestart Kornilowicz\surnameend} \&
  \bibinfo{author}{Adam \surnamestart Naumowicz\surnameend}
  (\bibinfo{year}{2010}): \emph{\bibinfo{title}{Mizar in a Nutshell}}.
\newblock {\sl \bibinfo{journal}{J. Formalized Reasoning}}
  \bibinfo{volume}{3}(\bibinfo{number}{2}), pp. \bibinfo{pages}{153--245},
  \doi{10.6092/issn.1972-5787/1980}.

\bibitemdeclare{inproceedings}{DBLP:conf/types/HaftmannW06}
\bibitem{DBLP:conf/types/HaftmannW06}
\bibinfo{author}{Florian \surnamestart Haftmann\surnameend} \&
  \bibinfo{author}{Makarius \surnamestart Wenzel\surnameend}
  (\bibinfo{year}{2006}): \emph{\bibinfo{title}{Constructive Type Classes in
  Isabelle}}.
\newblock In \bibinfo{editor}{Thorsten \surnamestart Altenkirch\surnameend} \&
  \bibinfo{editor}{Conor \surnamestart McBride\surnameend}, editors: {\sl
  \bibinfo{booktitle}{Types for Proofs and Programs, International Workshop,
  {TYPES} 2006, Nottingham, UK, April 18-21, 2006, Revised Selected Papers}},
  {\sl \bibinfo{series}{Lecture Notes in Computer Science}}
  \bibinfo{volume}{4502}, \bibinfo{publisher}{Springer}, pp.
  \bibinfo{pages}{160--174}, \doi{10.1007/978-3-540-74464-1\_11}.

\bibitemdeclare{inproceedings}{KuArMyOw14}
\bibitem{KuArMyOw14}
\bibinfo{author}{Ramana \surnamestart Kumar\surnameend}, \bibinfo{author}{Rob
  \surnamestart Arthan\surnameend}, \bibinfo{author}{Magnus~O. \surnamestart
  Myreen\surnameend} \& \bibinfo{author}{Scott \surnamestart Owens\surnameend}
  (\bibinfo{year}{2014}): \emph{\bibinfo{title}{{{HOL}} with {{Definitions}}:
  {{Semantics}}, {{Soundness}}, and a {{Verified Implementation}}}}.
\newblock In: {\sl \bibinfo{booktitle}{Interactive {{Theorem Proving}}}},
  \bibinfo{publisher}{{Springer, Cham}}, pp. \bibinfo{pages}{308--324},
  \doi{10.1007/978-3-319-08970-6_20}.

\bibitemdeclare{article}{DBLP:journals/jar/KumarAMO16}
\bibitem{DBLP:journals/jar/KumarAMO16}
\bibinfo{author}{Ramana \surnamestart Kumar\surnameend}, \bibinfo{author}{Rob
  \surnamestart Arthan\surnameend}, \bibinfo{author}{Magnus~O. \surnamestart
  Myreen\surnameend} \& \bibinfo{author}{Scott \surnamestart Owens\surnameend}
  (\bibinfo{year}{2016}): \emph{\bibinfo{title}{Self-Formalisation of
  Higher-Order Logic - Semantics, Soundness, and a Verified Implementation}}.
\newblock {\sl \bibinfo{journal}{J. Autom. Reasoning}}
  \bibinfo{volume}{56}(\bibinfo{number}{3}), \doi{10.1007/s10817-015-9357-x}.

\bibitemdeclare{inproceedings}{DBLP:conf/cpp/Kuncar15}
\bibitem{DBLP:conf/cpp/Kuncar15}
\bibinfo{author}{Ondrej \surnamestart Kuncar\surnameend}
  (\bibinfo{year}{2015}): \emph{\bibinfo{title}{Correctness of Isabelle's
  Cyclicity Checker: Implementability of Overloading in Proof Assistants}}.
\newblock In \bibinfo{editor}{Xavier \surnamestart Leroy\surnameend} \&
  \bibinfo{editor}{Alwen \surnamestart Tiu\surnameend}, editors: {\sl
  \bibinfo{booktitle}{Proceedings of the 2015 Conference on Certified Programs
  and Proofs, {CPP} 2015, Mumbai, India, January 15-17, 2015}},
  \bibinfo{publisher}{{ACM}}, pp. \bibinfo{pages}{85--94},
  \doi{10.1145/2676724.2693175}.

\bibitemdeclare{article}{DBLP:journals/pacmpl/Kuncar018}
\bibitem{DBLP:journals/pacmpl/Kuncar018}
\bibinfo{author}{Ondrej \surnamestart Kuncar\surnameend} \&
  \bibinfo{author}{Andrei \surnamestart Popescu\surnameend}
  (\bibinfo{year}{2018}): \emph{\bibinfo{title}{Safety and conservativity of
  definitions in {HOL} and Isabelle/HOL}}.
\newblock {\sl \bibinfo{journal}{{PACMPL}}}
  \bibinfo{volume}{2}(\bibinfo{number}{{POPL}}), pp.
  \bibinfo{pages}{24:1--24:26}, \doi{10.1145/3158112}.

\bibitemdeclare{article}{DBLP:journals/jar/KuncarP19a}
\bibitem{DBLP:journals/jar/KuncarP19a}
\bibinfo{author}{Ondrej \surnamestart Kuncar\surnameend} \&
  \bibinfo{author}{Andrei \surnamestart Popescu\surnameend}
  (\bibinfo{year}{2019}): \emph{\bibinfo{title}{A Consistent Foundation for
  Isabelle/HOL}}.
\newblock {\sl \bibinfo{journal}{J. Autom. Reasoning}}
  \bibinfo{volume}{62}(\bibinfo{number}{4}), pp. \bibinfo{pages}{531--555},
  \doi{10.1007/s10817-018-9454-8}.

\bibitemdeclare{inproceedings}{KuncarPopescu2017a}
\bibitem{KuncarPopescu2017a}
\bibinfo{author}{Ondřej \surnamestart Kunčar\surnameend} \&
  \bibinfo{author}{Andrei \surnamestart Popescu\surnameend}
  (\bibinfo{year}{2017}): \emph{\bibinfo{title}{Comprehending
  {{Isabelle}}/{{HOL}}'s {{Consistency}}}}.
\newblock In \bibinfo{editor}{Hongseok \surnamestart Yang\surnameend}, editor:
  {\sl \bibinfo{booktitle}{Programming {{Languages}} and {{Systems}} - 26th
  {{European Symposium}} on {{Programming}}, {{ESOP}} 2017, {{Held}} as
  {{Part}} of the {{European Joint Conferences}} on {{Theory}} and {{Practice}}
  of {{Software}}, {{ETAPS}} 2017, {{Uppsala}}, {{Sweden}}, {{April}} 22-29,
  2017, {{Proceedings}}}}, {\sl \bibinfo{series}{Lecture Notes in Computer
  Science}} \bibinfo{volume}{10201}, \bibinfo{publisher}{{Springer}}, pp.
  \bibinfo{pages}{724--749}, \doi{10.1007/978-3-662-54434-1_27}.

\bibitemdeclare{misc}{HOL14}
\bibitem{HOL14}
\bibinfo{author}{Michael \surnamestart Norrish\surnameend} \&
  \bibinfo{author}{Konrad \surnamestart Slind\surnameend}
  (\bibinfo{year}{2014}): \emph{\bibinfo{title}{The {{HOL System LOGIC}}}}.
\newblock
  \urlprefix\url{http://downloads.sourceforge.net/project/hol/hol/kananaskis-10/kananaskis-10-logic.pdf}.

\bibitemdeclare{inproceedings}{DBLP:conf/rta/Obua06}
\bibitem{DBLP:conf/rta/Obua06}
\bibinfo{author}{Steven \surnamestart Obua\surnameend} (\bibinfo{year}{2006}):
  \emph{\bibinfo{title}{Checking Conservativity of Overloaded Definitions in
  Higher-Order Logic}}.
\newblock In \bibinfo{editor}{Frank \surnamestart Pfenning\surnameend}, editor:
  {\sl \bibinfo{booktitle}{Term Rewriting and Applications, 17th International
  Conference, {RTA} 2006, Seattle, WA, USA, August 12-14, 2006, Proceedings}},
  {\sl \bibinfo{series}{Lecture Notes in Computer Science}}
  \bibinfo{volume}{4098}, \bibinfo{publisher}{Springer}, pp.
  \bibinfo{pages}{212--226}, \doi{10.1007/11805618\_16}.

\bibitemdeclare{incollection}{pitts1993}
\bibitem{pitts1993}
\bibinfo{author}{Andrew~M. \surnamestart Pitts\surnameend}
  (\bibinfo{year}{1993}): \emph{\bibinfo{title}{The {HOL} Logic}}.
\newblock In \bibinfo{editor}{M.J.C. \surnamestart Gordon\surnameend} \&
  \bibinfo{editor}{Tom \surnamestart Melham\surnameend}, editors: {\sl
  \bibinfo{booktitle}{Introduction to {HOL}: A Theorem-Proving Environment for
  Higher-Order Logic}}, \bibinfo{publisher}{Cambridge University Press}, pp.
  \bibinfo{pages}{191--232}.

\bibitemdeclare{inproceedings}{LPAR23:AmPoGen20}
\bibitem{LPAR23:AmPoGen20}
\bibinfo{author}{Johannes~{\AA}man \surnamestart Pohjola\surnameend} \&
  \bibinfo{author}{Arve \surnamestart Gengelbach\surnameend}
  (\bibinfo{year}{2020}): \emph{\bibinfo{title}{A Mechanised Semantics for HOL
  with Ad-hoc Overloading}}.
\newblock In \bibinfo{editor}{Elvira \surnamestart Albert\surnameend} \&
  \bibinfo{editor}{Laura \surnamestart Kov{\'{a}}cs\surnameend}, editors: {\sl
  \bibinfo{booktitle}{LPAR23. LPAR-23: 23rd International Conference on Logic
  for Programming, Artificial Intelligence and Reasoning}}, {\sl
  \bibinfo{series}{EPiC Series in Computing}}~\bibinfo{volume}{73},
  \bibinfo{publisher}{EasyChair}, pp. \bibinfo{pages}{498--515},
  \doi{10.29007/413d}.
\newblock \urlprefix\url{https://easychair.org/publications/paper/9Hcd}.

\bibitemdeclare{inproceedings}{Rudnicki92}
\bibitem{Rudnicki92}
\bibinfo{author}{Piotr \surnamestart Rudnicki\surnameend}
  (\bibinfo{year}{1992}): \emph{\bibinfo{title}{An Overview of the Mizar
  Project}}.
\newblock In \bibinfo{editor}{Bengt \surnamestart Nordström\surnameend},
  \bibinfo{editor}{Kent \surnamestart Petersson\surnameend} \&
  \bibinfo{editor}{Gordon \surnamestart Plotkin\surnameend}, editors: {\sl
  \bibinfo{booktitle}{Proceedings of the 1992 Workshop on Types for Proofs and
  Programs}}, pp. \bibinfo{pages}{311--332}.
\newblock \urlprefix\url{http://mizar.org/project/MizarOverview.pdf}.

\bibitemdeclare{book}{Sho67}
\bibitem{Sho67}
\bibinfo{author}{Joseph~R. \surnamestart Shoenfield\surnameend}
  (\bibinfo{year}{1967}): \emph{\bibinfo{title}{Mathematical {{Logic}}}}.
\newblock \bibinfo{publisher}{{A.K. Peters}}, \bibinfo{address}{Natick, Mass}.

\bibitemdeclare{inproceedings}{DBLP:conf/tphol/Wenzel97}
\bibitem{DBLP:conf/tphol/Wenzel97}
\bibinfo{author}{Markus \surnamestart Wenzel\surnameend}
  (\bibinfo{year}{1997}): \emph{\bibinfo{title}{Type Classes and Overloading in
  Higher-Order Logic}}.
\newblock In \bibinfo{editor}{Elsa~L. \surnamestart Gunter\surnameend} \&
  \bibinfo{editor}{Amy~P. \surnamestart Felty\surnameend}, editors: {\sl
  \bibinfo{booktitle}{Theorem Proving in Higher Order Logics, 10th
  International Conference, TPHOLs'97, Murray Hill, NJ, USA, August 19-22,
  1997, Proceedings}}, {\sl \bibinfo{series}{Lecture Notes in Computer
  Science}} \bibinfo{volume}{1275}, \bibinfo{publisher}{Springer}, pp.
  \bibinfo{pages}{307--322}, \doi{10.1007/BFb0028402}.

\end{thebibliography}

\end{document}